\documentclass[preprint2]{proto}
\usepackage{times}   

\newcommand{\refs}{\par\noindent\hangindent=1pc\hangafter=1}
\def\lesssim{\mathrel{\hbox{\rlap{\hbox{\lower4pt\hbox{$\sim$}}}\hbox{$<$}}}}
\def\gtrsim{\mathrel{\hbox{\rlap{\hbox{\lower4pt\hbox{$\sim$}}}\hbox{$>$}}}}

\begin{document}

\title{\textbf{\LARGE 
               Current Advances in the Methodology and Computational Simulation
               of the Formation of Low-Mass Stars}}
\author {\textbf{\large Richard I. Klein}}
\affil{\small\em University of California, Berkeley \\
and Lawrence Livermore National Laboratory}

\author {\textbf{\large Shu-ichiro Inutsuka}}
\affil{\small\em Kyoto University}

\author {\textbf{\large Paolo Padoan}}
\affil{\small\em University of California, San Diego}

\author {\textbf{\large Kohji Tomisaka}}
\affil{\small\em National Astronomical Observatory, Japan
}

\begin{abstract}
\baselineskip = 11pt
\leftskip = 0.65in 
\rightskip = 0.65in
\parindent=1pc
{\small 
Developing a theory of low-mass star formation ($\sim 0.1$ to 3~M$_{\odot}$) 
remains one of the most elusive and important goals of theoretical
astrophysics. The star-formation process is the outcome of the complex
dynamics of interstellar gas involving non-linear interactions of
turbulence, gravity, magnetic field and radiation. The evolution of
protostellar condensations, from the moment they are assembled by 
turbulent flows to the time they reach stellar densities, spans 
an enormous range of scales, resulting in a major computational 
challenge for simulations. Since the previous Protostars and Planets conference,
dramatic advances in the development of new numerical algorithmic
techniques have been successfully implemented on large scale parallel 
supercomputers. Among such techniques, Adaptive Mesh Refinement 
and Smooth Particle Hydrodynamics have provided frameworks to 
simulate the process of low-mass star formation with a very large
dynamic range. It is now feasible to explore the turbulent fragmentation 
of molecular clouds and the gravitational collapse of cores into stars
self-consistently within the same calculation. The increased
sophistication of these powerful methods comes with substantial caveats
associated with the use of the techniques and the interpretation of 
the numerical results. In this review, we examine what has been 
accomplished in the field and present a critique of both numerical
methods and scientific results. We stress that computational simulations
should obey the available observational constraints and demonstrate
numerical convergence. Failing this, results of numerical simulations
do not advance our understanding of low-mass star formation. 
 \\~\\~\\~}

\end{abstract}

\section{\textbf{INTRODUCTION}}

Most of the stars in the galaxy exist in gravitationally bound binary
and low-order multiple systems. Although several mechanisms have been
put forth to account for binary star formation, fragmentation has emerged
as the leading mechanism in the past decade ({\em Bodenheimer et al.}, 2000).
This point of view has been strengthened by observations that have shown 
the binary frequency among pre-main-sequence stars is comparable to or 
greater than that among nearby main-sequence stars ({\em Duchene et al.}, 1999).  
This suggests that most binary stars are formed during the protostellar collapse 
phase. Developing a theory for low mass star formation (~0.1 to 3 solar masses), 
which explains the physical properties of the formation of binary and multiple 
stellar systems, remains one of the most elusive and important goals of theoretical
astrophysics. 

Until very recently, the extreme variations in length scale 
inherent in the star formation process have made it difficult to perform 
accurate calculations of fragmentation and collapse, which are intrinsically 
three-dimensional in nature. Since the last review in {\em Protostars and Planets IV}
by {\em Bodenheimer et al.}, 2000, dramatic advances in the development of new 
numerical algorithmic techniques, including adaptive mesh refinement (AMR) and 
Smooth Particle Hydrodynamics (SPH), as well as advances in
large scale parallel machines, have allowed a significant
increase in the dynamic range of simulations of low mass-star formation.  
It is now feasible to explore the turbulent fragmentation 
of molecular clouds and the gravitational collapse of cores into stars
self-consistently within the same calculation. In this chapter we examine
what has been recently accomplished in the field of numerical simulation
of low-mass star formation, and we critically review both numerical methods and
scientific results.

\bigskip
\noindent
\textbf{1.1. Key Questions Posed by the Observations}
\bigskip

Observational surveys present us with a basic picture of star-forming regions 
including the structure and dynamics of star-forming clouds and the properties
of protostellar cores. A theory of star--formation should explain both the
large scale environment and the properties of protostellar cores self-consistently.

As shown first by {\em Larson} (1981), and later confirmed by many other studies,
star-forming regions are characterized by a correlation between internal 
velocity dispersion and size, $\delta V\approx 1 {\rm km/s} (L/1{\rm pc})^{0.4}$.
This scaling law has been interpreted as evidence of supersonic turbulence on
a wide range of scales. The turbulence can provide support against the
gravitational collapse, but can also create gravitationally unstable 
compressed regions through shocks. A theory of star formation should
elucidate whether turbulence controls the star formation rate and efficiency, 
or are those properties controlled by stellar outflows 
and winds? Are the scaling laws of turbulent flows
related to scaling laws of core and stellar properties? Is the turbulence setting
the initial density perturbations that collapse into stars? What is the 
effect of turbulence on the accretion of mass onto protostars?

On smaller scales, observational surveys have shown that prestellar cores
are elongated. Their density profiles are flat near the center, steeper at 
larger radii, may show very sharp edges and are sometimes consistent
with Bonnor-Ebert profiles (e.g., {\em Bacmann et al.}, 2000; {\em Alves et al.}, 
2001). Cores have rotational energies on the average only a few percent of their
binding energies (e.g., {\em Goodman et al.}, 1993). They are marginally 
supercritical ({\em Crutcher}, 1999) and their mass distribution
is very similar to the stellar mass distribution (e.g., {\em Motte et al.}, 1998, 
2001). The large majority of cores are found to contain stars 
(e.g., {\em Jijina et al.}, 1999),
and individual cores produce at most 2-3 protostars (e.g., {\em Goodwin and 
Kroupa}, 2005). A theory of low-mass star formation should be consistent with 
these observations and address: 
Why are prestellar cores so short-lived?
Why do they have Bonnor-Ebert profiles? How does the observed core angular momentum
affect the formation of binary and multiple systems? What is the role
played by magnetic fields in their evolution? Why is the mass 
distribution of prestellar cores so similar to the stellar initial mass 
function? Why are cores barely fragmenting into binaries or low multiplicity
systems?

Observations of young stellar populations provide important constraints as 
well. We know that young stars are always found in association with dense
gas with an efficiency 
$\sim 10$-20\%, much higher than the overall star formation
efficiency in GMCs, $\sim 1$-3\% (e.g., {\em Myers et al.}, 1986).
Stars are often found in clusters, of size ranging from 10 to 1000 members
(e.g., {\em Lada and Lada}, 2003). The stellar initial mass function
peaks around a fraction of a solar mass and its lognormal shape around
the peak is roughly the same in open clusters, globular clusters and 
field stars ({\em Chabrier}, 2003). What determines the efficiency of star 
formation? Why is the stellar multiplicity higher in younger populations? 
What determines the typical stellar mass and the initial mass function? 

Although answering all these questions is outside the scope of this review, we 
pose them because these are the questions to be addressed by the computational 
simulations of the formation of low-mass stars.

\bigskip
\noindent
\textbf{1.2. Generation of Initial Conditions Consistent with the Observations}
\bigskip

Simulations of low mass star formation should generate initial conditions
for the collapse of protostars consistent with the observed physical properties
of star forming clouds. This can be achieved if a relatively large scale
is simulated ($>1$~pc) with numerical methods that can accurately reproduce
fundamental statistics measured in molecular clouds. Such statistics include
i) scaling laws of velocity, density and magnetic fields; ii) mean relative 
values of turbulent, thermal, magnetic and gravitational energies (the 
normalization of the scaling laws).

{\em 1.2.1. Scaling Laws.} {\em Larson} (1981) found that velocity and size of interstellar 
clouds are correlated over many orders of magnitude in size. This correlation
has been confirmed by many more recent studies (e.g., {\em Fuller and Myers}, 1992; 
{\em Falgarone et al.}, 1992). The most accepted interpretation is that
the scaling law reflects the presence of supersonic turbulence in the ISM
(e.g., {\em Larson}, 1981; {\em Ossenkopf and Mac-Low}, 2002; 
{\em Heyer and Brunt}, 2004). Large scale velocity-column density correlations 
from molecular line surveys of giant molecular clouds also suggest a turbulent 
origin of the observed density enhancements ({\em Padoan et al.}, 2001).
Starting with the work of {\em Troland and Heiles} (1986), a correlation between 
magnetic field strength and gas density, $B\propto n^{1/2}$, has been reported for mean 
densities larger than $n\sim 100$~cm$^{-3}$. However, density and magnetic field 
scalings are very uncertain because both quantities are difficult to measure.  

{\em 1.2.2. Mean Energies.} Assuming an average temperature of $T=10$~K, Larson's 
velocity-size correlation corresponds to an rms sonic Mach number 
$M_{\rm s}\approx 5$ on the scale of 1~pc, and $M_{\rm s}\approx 1$ at 0.02~pc. 
So, on the average, the turbulent kinetic energy is larger than the thermal energy. 
Indirect evidence of super-Alfv\'{e}nic dynamics in giant molecular clouds has 
been presented by {\em Padoan and Nordlund} (1999) and {\em Padoan et al.} (2004a). 
They have shown that the magnetic energy averaged over a large scale has an 
intermediate value between the thermal and the kinetic energies, even if it 
can be significantly larger than this average value within dense prestellar 
cores. Observations suggest that in dense cores gravitational, kinetic, thermal 
and magnetic energies are all comparable. However, the magnetic energy is very 
difficult to estimate. Accounting for both detections and upper limits, there is 
a large dispersion in the ratio of magnetic to gravitational energy of dense cores 
({\em Crutcher et al.}, 1993; {\em Crutcher et al.}, 1999; {\em Bourke et al.}, 2001; 
{\em Nutter et al.}, 2004). In the case of turbulence that is super-Alfv\'{e}nic 
on the large scale, this dispersion and the $B$-$n$ relation are predicted to be 
real ({\em Padoan and Nordlund}, 1999).

In summary, large scale simulations may provide realistic 
initial and boundary conditions for protostellar collapse, but they must be 
consistent with the turbulent nature of the ISM. On the scale of giant molecular 
clouds, observations suggest the turbulence is on the average supersonic and 
super-Alfv\'{e}nic and its kinetic energy is roughly equal to the cloud
gravitational energy.

\bigskip
\textbf{2. A BRIEF SURVEY OF LOW-MASS STAR FORMATION MECHANISMS}
\bigskip

Although much progress in numerical simulations of collapse and 
fragmentation has been made in the intervening 6 years since PPIV, 
a self-consistent theory of binary and multiple star formation
that addresses the key questions posed by observations is still not 
at hand. As discussed by {\em Bodenheimer et al.}, 2000, binary and multiple 
formation can occur through the processes of (i) capture, (ii) fission,
(iii) prompt initial fragmentation, (iv) disk fragmentation and (v)
fragmentation during the protostellar collapse phase. 

A recent mechanism for multiple star formation has been put forth by {\em
Shu et al.}, (2000) and {\em Galli et al.}, (2001). They develop equilibrium 
models of strongly magnetized isopedic disks and explored their bifurcation 
to non-axi-symmetric, multi-lobed structures of increasing rotation rates.  
Possible problems with this mechanism include the observed low rotational
energies, the observed random alignment of disks with the ambient magnetic 
fields, the complexity of star-forming regions relative to the two dimensional
geometry and absence of turbulence in the model.

Disk fragmentation from gravitational instability can result in multiple
systems in an equilibrium disk if the minimum Toomre $Q$ parameter
falls below $\approx$ 1. However, {\em Bodenheimer et al.}, 2000, 
have pointed out that the required initial conditions to obtain $Q<1$
may not be easily realized since the mass accretion timescale is
significantly longer than the dynamical timescale throughout most of 
the evolution of the protostar. Disk fragmentation plays a key
role in one of the theories of the formation of Brown Dwarfs (BDs).  
This scenario, known as the "failed embryo" scenario, begins with a 
gravitationally unstable disk surrounding a protostar. The disk
fragments into a number of substellar objects. If the crossing time of 
the cluster of embryos is much less than the free-fall time of the
collapsing core, one or more of the members will be rapidly ejected
resulting in a BD ({\em Reipurth and Clarke}, 2001). 
Problems with this model include observational evidence of
BD clustering ({\em Duch$\hat{e}$ne et al.}, 2004) , Ly-$\alpha$
signatures of BD accretion (e.g., {\em Jayawardhana et al.}, 2002; 
{\em Natta et al.}, 2004; {\em Barrado y Navascu{\'e}s et al.}, 2004; 
{\em Mohanty et al.}, 2005) and evidence that individual cores produce 
only 2 or 3 stars ({\em Goodwin and Kroupa}, 2005).

Currently, there are two dominant models to explain what determines
the mass of stars. The Direct Gravitational Collapse theory suggests that 
star-forming turbulent clouds fragment into cores that eventually 
collapse to make individual stars or small multiple systems 
({\em Shu et al.}, 1987; {\em Padoan and Nordlund}, 2002, 2004). 
In contrast, the Competitive Accretion theory suggests that at birth
all stars are much smaller than the typical stellar mass and that the final
stellar mass is determined by the subsequent accretion of unbound gas from 
the clump ({\em Bonnell et al.},1998; {\em Bonnell et al.}, 2001; 
{\em Bate et al.}, 2005). Significant problems
with competitive accretion include the large value of the 
observed virial parameter relative to the one required by competitive 
accretion ({\em Krumholz et al.}, 2005b). We discuss this problem
with competitive accretion in detail in section 4d.

\bigskip
\textbf{3. PHYSICAL PROCESSES NECESSARY FOR DETAILED SIMULATION OF LOW-MASS STAR FORMATION}
\bigskip

\noindent
\textbf{3.1. Turbulence}
\bigskip

The Reynolds number estimates the relative importance of the nonlinear
advection term and the viscosity term in the Navier-Stokes equation,
$Re=V_0L_0/\nu$. $V_0$ is the flow rms velocity, $L_0$ is the 
integral scale of the turbulence (say the energy injection scale)
and $\nu$ is the kinematic viscosity that we can approximate
as $\nu\approx v_{th}/(\sigma n)$. $v_{th}$ is the gas thermal 
velocity, $n$ is the gas mean number density and 
$\sigma\sim 10^{-15}$~cm$^2$ is the typical gas collisional cross section.
For typical molecular cloud values, $Re\sim 10^7$-$10^8$, which implies
flows are highly unstable to turbulence. It is important to recognize
the significance of turbulent gas dynamics in astrophysical 
processes, as turbulence is a dominant transport mechanism.
In molecular clouds, the turbulence is supersonic and the 
postshock gas cooling time is very short. This results in the 
highly fragmented structure of molecular clouds.

There has been significant progress in our understanding of 
supersonic turbulence in recent years (progress on the scaling 
properties of subsonic and sub-Alfv\'{e}nic turbulence is reviewed 
in {\em Cho et al.}, 2003). Phenomenological 
models of the intermittency of incompressible turbulence 
(e.g., {\em She and Leveque}, 1994; {\em Dubrulle}, 1994) have been extended 
to supersonic turbulence by {\em Boldyrev} (2002) and the predictions of 
the model have been confirmed by numerical simulations 
({\em Boldyrev et al.}, 2002a; {\em Padoan et al.}, 2004b). 
The intermittency correction is small for the exponent
of the velocity power spectrum (corresponding to the second order
velocity structure function) and large only at high order. 
However, {\em Boldyrev et al.} (2002b) have shown that low 
order density correlators depend on high order velocity 
statistics, so intermittency is likely to play a significant role 
in turbulent fragmentation, despite being only a small effect in the 
velocity power spectrum. 

Because supersonic turbulence can naturally generate, at very small scale,
strong density enhancements of mass comparable to a low mass stars
or even a brown dwarf, its correct description is of paramount importance 
for simulations of molecular cloud fragmentation into low mass 
stars and brown dwarfs. At present, the largest simulations of 
supersonic turbulence may achieve a Reynolds number $Re\sim 10^4$. 
The scale of turbulence dissipation is therefore much larger in 
numerical simulations (of order the computational mesh size) than 
in nature ($\sim 10^{14}$~cm). However, the ratio of the Kolmogorov 
dissipation scale,$\eta_{\rm K}$, and the Jeans length, $\lambda_{\rm J}$, 
is very small and remarkably independent of temperature and density, 
$\eta_{\rm K}/\lambda_{\rm J}\approx 10^{-4}
(T/10{\rm K})^{-1/8}(n/10^3{\rm cm}^{-3})^{-1/4}$.
One may hope to successfully simulate the process of turbulent fragmentation 
by numerically resolving the turbulence to scales smaller than $\lambda_{\rm J}$, 
but not as small as $\eta_{\rm K}$, unless the nature of turbulent 
flows varies dramatically between $Re\sim 10^3$ and $Re\sim 10^7$.
Experimental results seem to indicate that the asymptotic behavior 
of turbulence is recovered in the approximate range $Re=10,000$--20,000 
({\em Dimotakis}, 2000), which can be achieved with PPM simulations 
on a $2,048^3$ mesh.

In order to i) generate a sizable inertial range
(a power law power spectrum of the turbulent velocity over
an extended range of scales) and ii) resolve the turbulence just below
the Jeans length, a minimum computational box size of at least $1,000^3$ 
zones is required for a grid code. This accounts for the fact
that the velocity power spectrum starts to decay with increasing wavenumber 
faster than a power law already at approximately 30 
times the Nyquist frequency. It is still an optimistic estimate,
because at this resolution the velocity power spectrum is also
affected by the bottleneck effect (e.g., {\em Falkovich}, 1994; 
{\em Dobler et al.}, 2003; {\em Haugen and Brandenburg}, 2004).
Assuming the standard SPH kernel of 50 particles, this corresponds to 
at least $50\times1,000^3$ particles to describe the density field, 
and at least $few \times 1,000^3$ particles to describe the velocity 
field, if a Godunov SPH method is used (see below). 

Grid code simulations have started to achieve this dynamical range
only recently, while particle codes appear unsuitable to the task.
The calculation of {\em Bate et al.} (2003) has $3.5\times 100^3$ 
particles, more than four orders of magnitude below the above estimate
and therefore inadequate to describe the process of turbulent 
fragmentation (the formation of small scale density enhancements by
the supersonic turbulence). Studies proposing to directly test the 
effect of turbulence on star formation, based on numerical simulations
with resolution well below the above estimate, should be regarded with 
caution.

\bigskip
\noindent
\textbf{3.2. Gravity}
\bigskip

\noindent
{\em 3.2.1. The Jeans Condition.}
{\em Jeans} (1902) analyzed the linearized equations of
1D isothermal self-gravitational hydrodynamics (GHD)
for a medium of infinite extent and found that
perturbations on scales larger than the Jeans length,
$\lambda_J\equiv\ (\frac{\pi{c_s}^2}{G\rho})^{1/2}$,
are unstable. Thermal pressure cannot resist the self-gravity of
a perturbation larger than $ \lambda_J$, resulting in runaway collapse.
{\em Truelove et al.} (1997) showed that the errors
generated by numerical GHD solvers can act as
unstable perturbations to the flow. In a simulation with 
variable resolution, cell-scale errors introduced in
regions of coarser resolution can be advected to regions of
finer resolution, allowing these errors to grow.
The unstable collapse of numerical perturbations
can lead to artificial fragmentation. The strategy to avoid
artificial fragmentation is to maintain a sufficient
resolution of $\lambda_J$. Defining the Jeans number
$J\equiv\frac{\Delta x}{\lambda_J}$, {\em Truelove et al.}
(1997) found that keeping $J\leq0.25$ avoided
artificial fragmentation in the isothermal evolution of a
collapse spanning 7 decades of density, the approximate
range separating typical molecular cloud cores from 
nonisothermal protostellar fragments. This Jeans condition 
arises because perturbations on scales above $\lambda_J$ are 
physically unstable, and discretization of the GHD PDEs introduces 
perturbations on all scales above $\Delta x$. It is essential to 
keep $\lambda_J$ as resolved as possible in order to diminish 
the initial amplitude of perturbations that exceed this scale.
Although it has been shown to hold only for isothermal evolution, 
it is reasonable to expect that it is necessary (although not necessarily 
sufficient) for nonisothermal collapse as well where the transition 
to non-isothermal evolution may produce structure on smaller scales 
than the local Jeans length. .

\noindent
{\em 3.2.2. Runaway Collapse.}
The self-gravitational collapse in nearly spherical 
geometry tends to show a so-called ``runaway collapse,'',
where the denser central region collapses much faster than the 
less-dense surrounding region. The mass of the central fast 
collapsing region is of the order of the Jeans mass, 
$M_{\rm J} = \rho \lambda_{\rm J}^3 \sim 
G^{-3/2} C_s^{3/2} \rho^{-1/2}$, 
which decreases monotonically in this runaway stage.  
The description of this process requires increasingly higher 
resolution, not only on the spatial scale but also on the mass scale.  
Therefore, an accurate description is not guaranteed even with 
Lagrangian particle methods such as SPH, if the number of particles is conserved. 
The end of the runaway stage corresponds to the deceleration of the 
gravitational collapse. If the effective ratio of specific heats, 
$\gamma$ ($P\propto \rho^\gamma$),  
becomes larger than $\gamma_{\rm crit} = 4/3$, 
the increased pressure can decelerate the gravitational collapse. 
For example, the question of how and when the isothermal evolution 
terminates was explored in {\em Masunaga and Inutsuka} (1999). 

\noindent
{\em 3.2.3. Thermal Budget.}
In the low density regime the gas temperature is affected by various 
heating and cooling processes (e.g., {\em Wolfire et al.}, 1995; 
{\em Koyama and Inutsuka}, 2000; {\em Juvela et al.}, 2001). 
However, above a gas density of $10^4$-$10^5$~cm$^{-3}$, depending on 
the timescale of interest, the gas is thermally well coupled with
the dust grains that maintain a temperature of order 10K. 
During the dynamical collapse, gas and dust are isothermal until 
a density of $10^{10}$-$10^{11}$~cm$^{-3}$, when the compressional 
heating rate becomes larger than the cooling rate ({\it Inutsuka and Miyama}, 
1997; {\it Masunaga and Inutsuka}, 1999). The further evolution of a 
collapsing core and the formation of a protostar are radiation-hydrodynamical 
(RHD) processes that should be modeled by solving the radiation transfer and the
hydrodynamics simultaneously and in three dimensions. Presently, the most 
sophisticated multi-dimensional models are based on the (flux-limited) diffusion 
approximation ({\em Bodenheimer et al.} 1990, {\em Krumholz et al.} 2005c). 

Once the compressional heating dominates the radiative cooling, 
the central temperature increases gradually from the initial value 
of $\sim$ 10 K. The initial slope of the temperature as a function
of gas density corresponds to an effective ratio of specific heats 
$\gamma=5/3$: $T(\rho)\propto \rho^{2/3}$ for $10 {\rm K}~<~T~<~100 {\rm K}$. 
This monatomic gas property is due to the fact that 
the rotational degree of freedom of molecular hydrogen 
is not excited in this low temperature regime (e.g., 
$E(J=2-0)/k_{\rm B} = 512 {\rm K}$). When the temperature becomes larger than 
$\sim 10^2 {\rm K}$, the slope corresponds to $\gamma=7/5$, as for diatomic 
molecules. This value of $\gamma$ is larger than the minimum required 
for thermal pressure support against gravitational collapse: 
$\gamma~>~\gamma_{\rm crit}~\equiv~4/3$. The collapse is therefore 
decelerated and a shock is formed at the surface of a quasi-adiabatic core, 
called ``the first core''. Its radius is about 1~AU in spherically symmetric 
models, but can be significantly larger in more realistic multi-dimensional 
models. It consists mainly of H$_2$. 

The increase of density and temperature inside the first core is 
slow but monotonic. When the temperature becomes $>~10^3~{\rm K}$, the 
dissociation of H$_2$ starts. The dissociation of H$_2$ acts as an 
efficient cooling of the gas, which makes $\gamma~<~4/3$, 
triggering the second dynamical collapse. In this second collapse phase, 
the collapsing velocity becomes very large and engulfs the first core.   
As a result, the first core lasts for only $\sim 10^3$ years. 
In the course of the second collapse, the central density attains 
the stellar value, $\rho \sim 1$~g/cm$^3$, and the second adiabatic core, 
or ``protostar'', is formed. The time evolution of the SED obtained from 
the self-consistent RHD calculation can be found in {\em Masunaga et al.} 
(1998) and {\em Masunaga and Inutsuka} (2000a,b). 

\noindent
{\em 3.2.4. Core Fragmentation.}
{\it Tsuribe and Inutsuka} (1999a,b) have shown that the 
fragmentation of a rotating collapsing core into a multiple system 
is difficult in the isothermal stage. {\it Matsumoto and Hanawa} 
(2003) have extended the collapse calculation by using a nested-grid 
hydro code and a simplified barotropic equation of state that mimics 
the thermal evolution, and have shown that the first-core disk increases 
the rotation-to-gravitational energy ratio ($T/|W|$) by mass accretion. 
A stability analysis of a rotating polytropic gas shows that gas with 
$T/|W| > 0.27$ is unstable for non-axisymmetric perturbations 
(e.g., {\it Imamura et al}. 2000). If the first-core disk rotates fast enough
that the angular speed $\times$ the free-fall time satisfies 
$\Omega_c (4\pi G\rho_c)^{-1/2} \gtrsim (0.2-0.3)$,
fragments appear and grow into binaries and multiples in the first core phase.
The non-axisymmetric nonlinear spiral pattern can transfer 
the angular momentum of the accreting gas.

\bigskip
\noindent
\textbf{3.3. Magnetic Fields}
\bigskip

Detailed self-consistent calculations accounting for both thermal and 
magnetic support ({\em Mouschovias and Spitzer}, 1976; {\em Tomisaka et al.}, 1988) 
show that the maximum stable mass can be expressed as
$M_\mathrm{mag,max}\sim M_\mathrm{BE}\left\{1-\left[0.17/(G^{1/2}M/\Phi)_{\mathrm c} \right]^2\right\}^{-3/2}$,
where $(M/\Phi)_{\mathrm c}$ is the central mass-to-flux ratio and 
$M_\mathrm{BE}=1.18c_{\mathrm s}^4/G^{3/2}p_{\mathrm {ext}}^{1/2}$
is the Bonnor-Ebert mass ({\em Bonnor}, 1956, 1957; {\em Ebert}, 1957).
A similar formula was proposed by {\em McKee} (1989), 
$M_\mathrm{mag,max}\sim M_\mathrm{BE}+\Phi_B/2\pi G^{1/2}$. 

Further support is provided by rotation. For a core with specific angular 
momentum $j$, the maximum stable mass is given by 
$M_\mathrm{max}\sim \left[M_\mathrm{mag,max}^2+(4.8 c_sj/G)^2\right]^{1/2}$
({\em Tomisaka et al.}, 1989). The dynamical runaway collapse begins when 
the core mass exceeds this maximum stable mass (magnetically supercritical 
cloud). Quasi-static equilibrium configurations exist for cores less massive than
the maximum stable mass. The evolution of these subcritical cores is 
controlled by the processes of ambipolar diffusion and magnetic braking, both 
of which have longer timescales than the gravitational free fall.
As the core contracts, the density grows and, when 
$n \gtrsim 10^{12} \mathrm{cm}^{-3}$,
the magnetic field is effectively decoupled from the gas. At these densities,  
Joule dissipation becomes important and particle drifts are qualitatively 
different from ambipolar diffusion ({\em Nakano et al.}, 2002).

The magnetic field is also responsible for the transfer of angular momentum in 
magnetized rotating cores, by a process called magnetic braking. 
Magnetic breaking is caused by the azimuthal component of the Lorentz force 
$(\vec{j}\times \vec{B})_\phi$. In the evolution of subcritical cores, the 
magnetic braking is important during the quasi-static contraction phase 
controlled by the ambipolar diffusion ({\em Basu and Mouschovias}, 1994).
In the dynamical runaway collapse, the rotational speed is smaller 
than the inflow speed ({\em Tomisaka}, 2000)

\bigskip
\noindent
\textbf{3.4. Outflows}
\bigskip

The magnetic field generates an outflow, by which star forming gas loses 
its angular momentum and accretes onto a protostar. Magneto-hydrodynamical 
simulations of the contraction of molecular cores ({\em Tomisaka}, 1998, 2000, 
2002; {\em Allen et al.}, 2003; {\em Banerjee and Pudritz}, 2006) have shown
that after the formation of the first core, the gas rotates around the core, 
a toroidal magnetic field is induced and magnetic torques transfer
angular momentum from the disk midplane to the surface. Outflows are accelerated 
in two ways: i) The gas which has received enough angular momentum compared with 
the gravity is ejected by the excess centrifugal force (magnetocentrifugal wind 
mechanism; {\em Blandford and Payne}, 1982). ii) In a core with a weak magnetic field, 
the magnetic pressure gradient of the toroidal magnetic field accelerates the 
gas and an outflow is formed in the direction perpendicular to the disk.  

Axisymmetric 2D simulations show that i) at least 10\% of the accreted mass is 
ejected; ii) the angular momentum is reduced to a factor $10^{-4}$ of the value of the 
parent cloud at the age of $\simeq 7000 {\rm yr}$ from the core formation. 
This resolves the angular momentum problem ({\em Tomisaka}, 2000). 
7000~yr from the first core formation, the mass of the core reaches 
$\sim 0.1 M_\odot$ and the outflow extends to a distance from the core of
$\simeq 2000 \mathrm{AU}$ with a speed of $\sim 2 \mathrm{km\,s}^{-1}$
({\em Tomisaka}, 2002). If the accretion continues and the core mass grows 
to one solar mass, the outflow expands and its speed is further increased.
It should be noted that the outflow refers to the physics of the first core 
collapse only; the energetics of outflows during the second core collapse phase
are yet to be determined.

\bigskip
\noindent
\textbf{3.5. Radiative Transfer in Multi-Dimensions}
\bigskip

Radiation transport has been shown to play a significant
role in the outcome of fragmentation into binary and multiple systems
({\em Boss et al.}, 2000; {\em Whitehouse and Bate}, 2005) 
and in limiting the largest stellar mass in 2D 
({\em Yorke and Sonnhalter}, 2002) and 3D simulations ({\em 
Edgar and Clarke}, 2004; {\em Krumholz et al.}, 
2005c). The strong dependence of the evolution of isothermal and nonisothermal
cloud models on the handling of the cloud's thermodynamics implies that
collapse calculations must treat the thermodynamics accurately in order to
obtain the correct solution ({\em Boss et al.}, 2000). Because of the 
great computational burden imposed by solving the mean intensity equation 
in the Eddington approximation (the computational time is increased by
a factor of 10 or more) it is tempting to sidestep the Eddington approximation 
solution altogether and employ a simple barotropic prescription
(e.g., {\em Boss}, 1981; {\em Bonnell}, 1994; {\em Bonnell and Bate},
1994a,; {\em Burkert et al.}, 1997; {\em Klein et al.}, 1999).  
However, {\em Boss et al.} (2000) showed that a simple barotropic 
approximation is insufficient and radiative transfer must be used.  
We discuss the various methods of radiation transport in section 3.b.6.

\bigskip
\textbf{4. METHODOLOGY OF NUMERICAL SIMULATIONS}
\bigskip

\noindent
\textbf{4.1. Complexity of the Problem of Low Mass Star Formation}
\bigskip

The computational challenge for simulations of low mass star formation 
is that star formation occurs in clouds over a huge dynamic range of spatial
scales, with different physical mechanisms being important on different
scales. The gas densities in these clouds also varies over many orders of 
magnitude as a result of supersonic turbulence and gravitational collapse. 
Gravity, turbulence, radiation 
and magnetic fields all contribute to the star formation process. Thus
the numerical problem is multi-scale, multi-physics and highly 
non-linear. To develop a feel for the range of scale a simulation must cover,
we can consider the internal structure of GMCs as hierarchical, consisting 
of smaller subunits within larger ones ({\em Elmegreen and Falgarone}, 1996). 
GMCs vary in size from 20 to 100~pc., in density from 50 to 100 H$_2$ cm$^{-3}$ and
in mass from $10^4$ to $10^6$ M$_{\odot}$.

Self-gravity and turbulence are equally important in controlling the
structure and evolution of these clouds. Magnetic fields are likely to play 
an important role as well ({\em Heiles et al.}, 1993; {\em McKee
et al.}, 1993). Embedded within the GMCs are dense clumps that may form 
clusters of stars. These clumps are few pc. in size, have masses of a few
thousand M$_{\odot}$ and mean densities $\sim 10^3$ H$_2$ cm$^{-3}$.  
The clumps contain dense cores with radii $\sim$ 0.1pc.,
densities $10^4$-$10^6$ H$_2$ cm$^{-3}$ and masses ranging from 1 to
several M$_{\odot}$. These cores likely form individual stars or low 
order multiple systems. The role of turbulence and magnetic fields in 
the fragmentation of molecular clouds has been investigated by 3D numerical
simulations (e.g., {\em Padoan and Nordlund}, 1999; {\em Ostriker et al.}, 
1999; {\em Ballesteros-Paredes}, 2003; {\em Mac Low and Klessen}, 2003; 
{\em Nordlund and Padoan}, 2003). 

A simulation that starts from a region of a turbulent 
molecular clouds (R $\sim$ few pc.) and evolves through the isothermal
core collapse into the formation of the first hydrostatic core at densities 
of $10^{13}$ H$_2$ cm$^{-3}$ requires an accurate calculation across
10 orders of magnitude in density and 4-5 orders of magnitude in spatial
scale. To resolve 100 AU separation binaries, one needs a resolution of
about 10 AU. To follow the collapse all the way to an actual star would 
require a further 10 orders of magnitude increase in density and 2-3  
more orders of magnitude in spatial scale. Such extraordinary 
computational demands rule out fixed grid simulations entirely  
and can be addressed only with accurate AMR or SPH approaches.

\bigskip
\noindent
\textbf{4.2. Smooth Particle Hydrodynamics}
\bigskip

The description of the gravitational collapse requires a large dynamic 
range of spatial resolution. An efficient way to achieve this is to use 
Lagrangian methods. Smoothed particle hydrodynamics (SPH) is a fully Lagrangian 
particle method designed to describe compressible fluid dynamics. This method 
is economical in handling hydrodynamic problems that have large, almost empty 
regions. A variety of astrophysical problems including star formation 
have been studied with SPH, because of its simplicity in programming  
two- and three-dimensional codes and its versatility to incorporate 
self-gravity. A broad discussion of the method can be found in a review by 
{\it Monaghan} (1994). An advantage of SPH is its conservation property;  
SPH is Galilean invariant and, in contrast to grid-based methods, conserves 
both linear and angular momentum simultaneously. The method to conserve the 
total energy within a computer round-off-error is explained in {\it Inutsuka} (2002). 
In order to further increase the dynamic range of spatial resolution, 
{\it Kitsionas and Whitworth} (2002) introduced particle splitting, which is 
an adaptive approach in SPH.  

The ``standard'' SPH formalism adopts artificial viscosity that mimics the 
classical von-Neumann Richtmeyer viscosity. This tends to give poor 
performance in the description of strong shocks. In two- or three-dimensional 
calculations of colliding streams, standard SPH particles often penetrate into 
the opposite side. This unphysical effect can be partially eliminated by 
the so called XSPH prescription ({\em Monaghan}, 1989), which does not introduce 
the (required) additional {\em dissipation}, but results in additional {\em dispersion} 
of the waves. As a more efficient method for handling strong shocks in the SPH 
framework, the so called ``Godunov SPH'' was proposed by {\it Inutsuka} (2002), 
who implemented the exact Riemann solver in the strictly conservative particle method. 
This was used in the simulation of the collapse and fragmentation of self-gravitating 
cores ({\it Tsuribe and Inutsuka}, 1999a; {\it Cha and Whitworth}, 2003a,b). 

The implementation of self-gravity in SPH is relatively easy and one can use 
various acceleration methods, such as {\em Tree-Codes}, and special purpose 
processors (e.g., GRAPE board).  
The flux-limited diffusion radiative transfer was recently incorporated in SPH 
by {\em Whitehouse and Bate} (2004), {\em Whitehouse et al.} (2005) and 
{\em Bastien, Cha, and Viau} (2004). 

Several groups are now using ``sink particles'' to follow the 
subsequent evolution even after protostars are formed 
({\em Bate et al.}, 1995). This is a prescription to continue the 
calculations without resolving processes of extremely short timescale 
around stellar objects. {\em Krumholz et al.}, (2004) have introduced sink 
particles for the first time into Eulerian grid-based methods and in 
particular for AMR.

\def\lesssim{\mathrel{\hbox{\rlap{\hbox{\lower4pt\hbox{$\sim$}}}\hbox{$<$}}}}
\def\gtrsim{\mathrel{\hbox{\rlap{\hbox{\lower4pt\hbox{$\sim$}}}\hbox{$>$}}}}
\newcommand{\ve}[1]{\mbox{\boldmath${#1}$}}

\bigskip
\noindent
\textbf{4.3. Fixed-Grid Hydrodynamics}
\bigskip

Since the time of its introduction, the numerical code of choice 
for supersonic hydrodynamic turbulence has been the Piecewise Parabolic
Method (http://www.lcse.umn.edu/) (PPM) of {\em Colella and Woodward} (1984). 
PPM is based on a Rieman solver (the discretized approximation to the solution 
is locally advanced analytically) with a third order accurate reconstruction scheme, 
which allows an accurate and stable treatment of strong shocks, while maintaining 
numerical viscosity to a minimum away from discontinuities. Because 
the physical viscosity is not explicitly computed (PPM solves the Euler equations),
large scale PPM flows are characterized by a very large effective 
Reynolds number ({\em Porter and Woodward}, 1994). Direct numerical
simulations (DNS) of the Navier-Stokes equation, where the physical 
viscosity is explicitly computed, require a linear numerical resolution four 
times larger than PPM to achieve the same wave-number extension of 
the inertial range of turbulence as PPM ({\em Sytine et al.}, 2000). From this
point of view, therefore, PPM has a significant advantage over DNS codes.
Furthermore, DNS codes are generally designed for incompressible turbulence, 
and hence of limited use for simulations of the ISM. 

Codes based on straightforward staggered mesh finite difference methods, rather than 
Rieman solvers, have also been used in applications to star formation and
interstellar turbulence, such as the 
Zeus code (http://cosmos.ucsd.edu/) 
({\em Stone and Norman}, 1992a,b) and the 
Stagger Code (www.astro.ku.dk/StaggerCode/) 
({\em Nordlund and Galsgaard}, 1995; {\em Gudiksen and Nordlund}, 2005). 
Finite difference codes address the problem of 
supersonic turbulence with the introduction of localized numerical 
viscosity to stabilize the shocks while keeping viscosity as low as 
possible away from shocks. The main advantage of this type of code, 
compared with Rieman solvers, is their flexibility in incorporating 
new physics and their computational efficiency. 

Fixed-grid codes cannot achieve the dynamical range required by problems 
involving the gravitational collapse of protostellar cores. Such problems are 
better addressed with particle methods such as SPH, or by generalizing the 
methods used for fixed-grid codes into AMR schemes. The main advantage of
large fixed-grid experiments is their ability to simulate the physics of
turbulent flows. As supersonic turbulence is believed to play a crucial role
in the initial fragmentation of star-forming clouds, fixed-grid codes
may still be the method of choice to generate realistic large scale initial 
conditions for the collapse of protostellar cores. Recent attempts of
simulating supersonic turbulence with AMR methods are promising ({\em Kritsuk et 
al.}, 2006), but may be truly advantageous only at a resolution above $\sim 1,000^3$.
SPH simulations to date have resolution far too small for the task, as commented above,
and have not been used so far as an alternative method to investigate the 
physics of turbulence.

\bigskip
\noindent
\textbf{4.4. Adaptive Mesh Refinement Hydrodynamics and Nested Grids}
\bigskip

The adaptive mesh refinement (AMR) scheme utilizes underlying rectangular 
grids at different levels of resolution. Linear resolution varies by integral
refinement factors between levels, and a given grid is
always fully contained within one at the next coarser level (excluding the
coarsest grid). The origin of the method stems from the seminal work of
{\em Berger and Oliger} (1984) and {\em Berger and Collela} (1989).  
The AMR method dynamically resizes and repositions
these grids and inserts new, finer ones within them according to
adjustable refinement criteria, such as the numerical Jean's condition 
({\em Truelove et al.}, 1997). Fine grids are automatically
removed as flow conditions require less resolution. During the
course of the calculation, some pointwise measure of the error is computed
at frequent intervals, typically every other time step. At those times,
the cells that are identified are covered by a relatively small number of
rectangular patches, which are refined by some even integer factor.

Refinement is in both time and space, so that the calculation on the refined 
grids is computed at the same Courant number as that on the coarse grid. 
The finite difference approximations 
on each level of refinement are in conservation form, 
as is the coupling at the interface between grids at different levels of 
refinement. AMR has three substantial advantages over standard SPH.  
Combined with high order Godunov methods, AMR achieves a much
higher resolution of shocks. This is important in obtaining accuracy in
supersonic turbulent flows in star forming clumps and cores and in accretion 
shocks onto forming protostars. AMR allows high resolution at {\em all} 
points in the flow as dictated by the physics. Unlike SPH, where
particles are taken away from low density regions, where accuracy is lost, 
and concentrated into high density regions, AMR maintains high
accuracy everywhere.  An important consequence of this is that if SPH
were to maintain the same comparable resolution as AMR everywhere 
in the flow, it would be prohibitively expensive. AMR is based 
on fixed Eulerian grids and thus can 
take advantage of sophisticated algorithms to incorporate magnetic fields 
and radiative transfer. This is far more difficult in a particle-based scheme.
AMR was first introduced into astrophysics by {\em Klein et al.} (1990, 1994) 
and has been used extensively both in low mass and high mass star
formation simulations ({\em Truelove et al.}, 1998;
{\em Klein}, 1999; {\em Klein et al.}, 2000, 2003, 2004a;
{\em Krumholz et al.}, 2005c).  

An advantage of SPH over {\em Cartesian} grid based AMR is that for pure 
hydrodynamics it can conserve both linear and angular momentum simultaneously 
to within round-off errors whereas Cartesian grid based AMR cannot. 
However, if one uses a cylindrical or spherical coordinate
system for the simulation of protostellar disks for instance, then grid based
AMR conserves total angular momentum to round-off.  We point out, however, that
these statements apply only to pure hydrodynamics.  Once forces such as gravity
are included, the situation becomes worse and both grid codes that solve the Poisson equation
or SPH codes that use tree-type acceleration or grid based methods for gravity
lose the conservation property for total linear momentum as well. 

There are several ways to implement AMR. They can be broadly divided into
two categories: Meshes with fixed number of cells, such as in Lagrangian
or rezoning approaches, and meshes with variable number of cells, such as
unstructured finite elements, structured cell-by-cell and structured
sub-grid blocks. For various reasons the most widely adopted approaches
in astrophysics are structured sub-grid blocks and cell-by-cell. 
The first was developed by {\em Berger and Oliger} (1984) 
and {\em Berger and Collela} (1989). It is used in the AMR code ORION 
developed by Klein and collaborators
({\em Klein}, 1999; {\em Crockett et al.}, 2005) and in the community
code ENZO ({\em Norman and Bryan}, 1999). 
The cell-by-cell approach  such as in PARAMESH ({MacNeice et al.}, 2000) is used
in the community code Flash ({\em Banerjee et al.}, 2004). A hybrid approach is
used in the code NIRVANA ({\em Ziegler}, 2005). The cell-by-cell method has 
the advantages of flexible and efficient
refinement patterns and low memory overhead and the disadvantages of expensive 
interpolation and derivation formulas and large tree data structures. The 
sub-grid block method is more efficient and more suitable for shock capturing 
schemes than the cell-by-cell method, at the price of some memory overhead.
 
Finally, nested grids consisting of concentric hierarchical 
rectangular subgrids can also be very effective for problems of well defined
geometry ({\em Yorke et al.}, 1993). These methods are advantageous for tracing 
the non-homologous runaway collapse of an initially symmetrical cloud in 
which the coordinates of a future dense region are known in advance 
({\em Tomisaka}, 1998). The finest subgrid is added dynamically when 
spatial resolution is needed as in AMR methods.

\bigskip
\noindent
\textbf{4.5 Approaches for Magneto-Hydrodynamics}
\bigskip

Since strong shocks often appear in the astrophysical phenomena,
a shock-capturing scheme is needed also in MHD.   
Upwind schemes based on the Riemann solver are used as the MHD engine.
Schemes well known in hydrosimulations,
such as Roe's approximate Riemann solver 
({\em Brio and Wu}, 1988; {\em Ryu and Jones}, 1995;
{\em Nakajima and Hanawa}, 1996), 
piecewise parabolic method (PPM; {\em Dai and Woodward}, 1994), 
are also applicable to MHD.

Special attention should be paid to guarantee $\mathrm{div}\vec{B}=0$ in MHD
simulations. To ensure that the divergence of Maxwell stress tensor
 $T_{ij}=-(1/4\pi)B_iB_j+(1/8\pi)B^2\delta_{ij}$ 
 gives the Lorentz force, the first term of right-hand side
 $\partial_j (B_iB_j)$ must be equal to $B_j\partial_j B_i$.
This requires $B_i\partial_j B_j=0$ and means that a fictitious force 
appears along the magnetic field if the condition of divergence-free is broken.
The divergence of the magnetic field amplifies the instability of the solution
even for a linear wave. Thus, it is necessary for the MHD scheme
to keep the divergence of the magnetic field zero within a round-off error
or at least small enough. This divergence-free nature should be satisfied 
for the boundaries of subgrids in AMR and nested grid schemes.        

One solution is based on ``constrained transport (CT)'' 
({\em Evans and Hawley} 1988), in which the staggered collocation of the 
components of magnetic field on the cell faces makes the numerical divergence 
vanish exactly. In the staggered collocation, the electric field 
$-\vec{v\times B}$ of the induction equation 
$\partial_t \vec{B}=\vec{\nabla \times (v\times B)}$
is evaluated on the edge of the cell-face and the line integral along 
the edge gives the time difference of a component of the magnetic field.
Note that the electric field on one edge appears twice to complete the
induction equation. To guarantee a vanishing divergence of the magnetic field,
CT requires the two evaluations to coincide with each other.

To utilize the Godunov-type Riemann solver in the context of CT, 
{\em Balsara and Spicer} (1999) proposed a scheme as follows:
(1) face-centered magnetic field is interpolated to the cell center;
(2) from the cell-centered variables, the numerical flux at the cell face 
is obtained using a Riemann solver; (3) the flux is interpolated to the 
edge of the cell-face and the electric field in the induction equation is 
obtained; (4) new face-centered magnetic field is obtained from the induction 
equation. Variants of this method are widely used [see also 
{\em Ryu et al.}, (1998) and {\em Ziegler}, (2004)]. 

Avoiding staggered collocation of the magnetic field requires divergence cleaning.
In this case, divergence cleaning is realized by replacing the magnetic field 
every step as $\vec{B}^{new}=\vec{B}-\vec{\nabla \Phi}$,
where $\vec{\nabla^2}\Phi=\mathrm{div}\vec{B}$ (Hodge projection),
or by solving a diffusion equation for div $\vec{B}$ as
$\partial_t\vec{B}=\eta \nabla(\nabla\cdot\vec{B})$.
The former is combined with pure Godunov-type Riemann solvers using only
cell-centered variables ({\em Ryu et al.}, 1995). 
{\em Crockett et al.} (2005) reported that
the divergence cleaning of the face-centered magnetic field appearing
in the numerical flux based on an unsplit, cell-centered Godunov scheme
improves its accuracy and stability.     

{\em Powell et al.} (1999) proposed a different formalism,
in which div $\vec{B}$ term is kept in the MHD equations as a source
(e.g., the Lorentz force $(\nabla \times \vec{B})\times \vec{B}/4\pi$ gives 
an extra term related to div $\vec{B}$ as $-\vec{B}\nabla\cdot \vec{B}$
besides the Maxwell stress tensor term $-T_{ij}$,
in the equation for momentum density.) In this formalism, 
div $\vec{B}$ is not amplified but advected along the flow.
Comparison between various methods is found in {\em T\'{o}th} (2000),
{\em Dedner et al.} (2002), {\em Balsara and Kim} (2004) and {\em Crockett et al.} 
(2005).

There have been attempts to solve the induction equation with SPH
methods (e.g., {\em Stellingwerf and Peterkin}, 1994; {\em Byleveld and Pongracic}, 
1996; {\em Price and Monaghan}, 2004a,b,c, 2005). A major obstacle is an instability
that develops when the momentum and energy equations are written in 
conservation form. As a result, the equations must be written in a way that
does not conserve momentum ({\em Phillips and Monaghan}, 1985; {\em Morris}, 1996),
which is a major concern for the accurate treatment of shocks. Results of recent 
tests of the state-of-the-art SPH MHD code ({\em Price and Monaghan}, 2004c)
appear to be rather poor even for very mild shocks, and we conclude that MHD 
with SPH is not yet viable for simulations.

\bigskip
\noindent
\textbf{4.6. Approaches for Radiation Transport}
\bigskip

Several levels of approximation of the radiation transport in star formation simulations
can be used and details of various methods can be found in {\em Mihalas and Mihalas}
(1984) and {\em Castor} (2004). Here we briefly describe these methods
and point out their strengths and weaknesses. Although modern simulations using 
radiative transfer are still at an early stage, we include methods that hold promise for
the future that will circumvent the weaknesses of more approximate 
approaches currently in place. 

The simplest improvement beyond a barotropic stiffened EOS
is the diffusion approximation which pertains to the limit in which radiation
can be treated as an ideal fluid with small corrections. The
approximation holds when the photon mean free path is small compared 
with other length scales. The combined energy equation for the gas and 
radiation results in an implicit non-linear diffusion equation for the
temperature. The weakness of the diffusion approximation
is that it is strictly applicable to optically thick regimes and performs
poorly in optically thin regions. This can be severe in optically thin 
regions of an inhomogeneous turbulent core or in the optically thin atmosphere
surrounding a developing protostar. 

The next level of approximation is the Eddington approximation
({\em Boss and Myhill}, 1992; {\em Boss et al.}, 2000). It can be shown
that the diffusion approximation leads directly to Eddington's approximation 
$P_{\nu}$ = $\frac{1}{3}E_{\nu}I$, where $P_{\nu}$ is the pressure tensor
moment of the specific intensity of radiation, $E_{\nu}$ is the scalar
energy density of radiation and $I_{\nu}$ is the
isotropic identity tensor. This approximation, coupled with dropping
the time dependent term in the 2nd moment equation of transfer results in
a combined parabolic 2nd order time dependent diffusion
equation for the energy density of the radiation field.
This formulation of the Eddington approximation is used in {\em Boss et
al.} (2000). The approximation results
in a loss of the finite propagation speed of light $c$ and a loss of the
radiation momentum density, thus there is an error in the total momentum budget.  
In optically thin regions, the radiation flux can increase
without limit.  As with all diffusion like methods, this approach also suffers
from shadow effects whereby radiation will tend to fill in
behind optically thick structures and may lead to unphysical heating. 

An alternative approach is the Flux Limited Diffusion (FLD),
that modifies the Eddington approximation, and
compensates for the errors made in dropping the time dependent flux term by
including a correction factor in the diffusion coefficient for the radiation 
flux. This correction factor, called a flux limiter,
is in general a tensor and has the property that the flux goes to the
diffusion limit at large optical depth and it correctly limits the flux 
to be no larger than $cE$ in the optically thin regime.
This improvement over the Eddington approximation has been used by 
{\em Klein et al.} (2004c) for the simulation of
both low mass and high mass star formation.  
The resulting sparse matrices introduced by the diffusion like terms
are solved by multi-grid iterative methods in an AMR framework.  
The flux-limiting correction can cause errors of order 20%
in the flux-limiter (or the flux), similar to the errors of the Eddington
approximation of 20\% in the Eddington factor at $\tau$ = 0 in the
Milne problem ({\em Castor} 2004). The FLD method's also suffer from shadow 
effects which can be severe. 

The next level of approximation, the variable Eddington tensor method,
removes many of the inaccuracies of the Eddington approximation and the
flux limiter modification. It was first formulated in multi--dimensions 
for astrophysical problems by {\em Dykema et al.} (1996). In essence,
if the precise ratio of the pressure tensor to the energy density were
included as an ad hoc multiplier in the Eddington approximation equations 
they would represent an exact closure of the system. The tensor
ratio is obtained iteratively from either an auxiliary solution of the
exact transport equation for the specific intensity or using an approximate 
analytic representation of the tensor. This method holds promise for future 
simulations, but has yet to be used in star formation. 

The final two approaches, which are highly accurate and deal with the 
angle dependent transport equation directly, are $S_N$ methods and Monte 
Carlo methods. They have not yet been developed for simulations in star
formation because the cost in 3D is prohibitive. The $S_N$ method is a
short characteristic method in which a bundle of rays is created at 
every mesh point and are extended in the upwind direction only as far as 
the next spatial cell. The main problem is in finding the efficient angle 
set to represent the radiation field in 2 or 3 dimensions ({\em Castor}, 2004).  
Finally, one might consider Monte Carlo methods to solve the transport equation.  
Although simple to implement (its great advantage), this method suffers from 
needing a vast number of operations per timestep to get accurate
statistics in following the particles used to track the radiation field.
Both of these  methods will avoid shadow effects and may be necessary
to accurately treat optically thick inhomogeneous structures that form in 
accretion flows onto protostars. 

Radiative transfer implementations have recently been developed also 
for SPH methods, based on the flux-limited diffusion  
({\em Whitehouse et al.}, 2005) or the Monte Carlo method
({\em Stamatello and Whitworth}, 2003, 2005).

\bigskip
\noindent
\textbf{4.7. Comparison of Computational Methods}
\bigskip

Based on the physical processes and numerical methodologies discussed in 
the previous sections we can compare numerical schemes according to their
ability to handle the following problems both accurately and efficiently: 
(a) turbulence, (b) strong shocks, (c) self-gravity, (d) magnetic fields, 
(e) radiative transfer.  

The standard SPH method has been successful with (c) and implementations of 
(e) have been recently developed in the flux-limited diffusion approximation 
and with a Monte Carlo method. It does not include (d), it is well known 
to be inadequate for (b) and has had virtually no applications to (a) to date. 
As any Lagrangian particle methods, SPH provides good resolution in 
high density regions, but very poor in low density ones.
The Godunov SPH method improves the standard SPH codes because of its ability 
to address (b), but does not provide a solution to (d) and is untested for (a)
as well. Although MHD is currently under development in SPH, results of standard
MHD tests with a state-of-the-art code show the need for significant
improvements even in the case of very mild shocks. 
Current applications of SPH should therefore be limited to non-MHD problems and 
the accuracy and performance of SPH with turbulent flows must be thoroughly tested.

In hydrodynamical problems, grid-based methods 
such as MUSCL ({\em van Leer}, 1979) and PPM ({\em Colella and Woodward}, 1984) 
use exact Riemann Solvers to construct the numerical fluxes and provide very 
accurate description in astrophysical flows with strong shocks (b). They have 
also been thoroughly tested with compressible turbulent flows, and MHD versions
have been developed that can address both (d) and (e). Traditional 
finite-difference grid-based schemes may still be useful, because the best of 
them can also accurately address (a), (b), (d) and (e),
at a lower cost of code development and computer resources.
Point (c) can also be efficiently dealt with by grid-based codes thanks 
to AMR methods. However, the development of AMR schemes that satisfy (c)
and at the same time (d) has begun only recently. These schemes exist and 
have been successfully tested, but it is unclear at present which approach
will provide the best trade off between accuracy and performance.

The constrained transport method appears to be the ideal one
to guarantee the $\nabla \cdot B = 0$ condition. Various schemes have 
been proposed even in the category of Godunov-type methods with a 
linearized Riemann Solver. An exact MHD Riemann Solver would be 
more adequate to handle strong shocks, but it is not available yet. 
In MHD we have to solve seven characteristics even in one-dimensional 
problems, which hinders an efficient construction of numerical fluxes 
based on the non-linear waves. Furthermore, the discovery of the 
existence of the MHD intermediate shocks ({\em Brio and Wu}, 1988) 
brings an additional difficulty in the categorization and prediction 
of the emerging non-linear waves. Among possible solutions,
a linearized Riemann Solver with artificial viscosity may still be
a useful option. 

The Godunov MHD code of {\em Crockett et al}, (2005) has been merged with 
the AMR RHD code of {\em Klein et al.} (2004a,b) into the first fully
developed AMR magneto-radiation-hydrodynamic code (ORION) to be used in 
simulations of star formation capable of addressing (a) through (e).

\bigskip
\textbf{5. RECENT SIMULATIONS AND CONFRONTATION WITH THE OBSERVATIONS}
\bigskip

\noindent
\textbf{5.1. Turbulent Fragmentation of Molecular Clouds}
\bigskip

The fragmentation of molecular clouds is the result of a complex
interaction of supersonic turbulence with gravity and magnetic fields.
Supersonic turbulent flows generate nonlinear density enhancements
through a complex network of interacting shocks. Some density 
enhancements are massive and dense enough to undergo gravitational
instability and collapse into stars. 
A fundamental problem with our understanding of star formation is that
the physics of turbulence is not fully understood. In order to investigate 
the process of turbulent fragmentation we must rely on large numerical 
simulations that barely resolve the scale-free nature of interstellar 
turbulent flows. If the scaling laws of turbulence play a role in the
star formation process, we cannot accurately test their effects numerically,
unless those scaling laws are well reproduced in the simulations. 
{\em Ballesteros-Paredes et al.} (2006) have tested the idea 
that the power spectrum of turbulence determines the Salpeter IMF 
({\em Padoan and Nordlund}, 2002). However, their simulations 
do not generate a turbulence inertial range, due to the combined effect
of low resolution and large numerical diffusivity. Their velocity power spectra 
do not show any extended power laws, and they even differ between their 
grid and SPH simulations. As a consequence, such numerical simulations 
fail to reproduce a scale free mass distribution of unstable cores.

Other recent SPH simulations have been used to compute the stellar 
mass distribution (e.g., {\em Bonnell et al.}, 2003; {\em Klessen}, 
2001) and to test the effect of turbulence on star formation 
({\em Delgado-Donate et al.}, 2004). Although such SPH simulations are
ideally suited to follow the collapse of individual objects due to their 
Lagrangian nature, their size is far too small to generate an inertial range
of turbulence, as discussed in section 3.a.1. 

Despite their limitations, numerical simulations carried out over the last 
few years have given us a good statistical picture of the process
of fragmentation of magnetized supersonic flows. The following are 
the most important results: i) The dissipation time of turbulence is almost 
independent of the magnetic field strength. The turbulence decays in 
approximately one dynamical time in both equipartition and super-Alfv\'{e}nic 
flows ({\em Padoan and Nordlund}, 1997; {\em MacLow et al.}, 1998; {\em Stone et 
al.}, 1998; {\em Padoan and Nordlund}, 1999; {\em Biskamp and Muller}, 2000). 
ii) The velocity power spectrum and structure functions are power laws over an 
inertial range of scales ({\em Boldyrev}, 2002; {\em Boldyrev et al.} 2002a,b; 
{\em Padoan et al.}, 2004b). In the limit of very large rms Mach number, 
the turbulent velocity power spectrum scales approximately as 
$u_{\rm k}^2\propto k^{-1.8}$ and the velocity structure functions follow 
the relative scaling predicted by {\em Boldyrev} (2002). For intermediate 
levels of compressibility, the scaling exponents depend on the rms Mach number 
({\em Padoan et al.}, 2004b). iii) The power spectrum of the gas density is a 
power law over an inertial range of scales. Its slope is a function of the rms 
Mach number of the flow and of the average magnetic field strength 
({\em Padoan et al.}, 2004b; {\em Beresnyak et al.}, 2005). 
iv) With an isothermal equation of state, 
the probability density function of gas density is well approximated by a 
Log-Normal distribution ({\em Vazquez-Semadeni}, 1994; {\em Padoan et al.},
1997; {\em Scalo et al.}, 1998; {\em Passot and Vazquez-Semadeni}, 1998; 
{\em Nordlund and Padoan}, 1999; {\em Ostriker et al.}, 1999; {\em Wada and Norman}, 
2001), with the dispersion of linear density proportional to the rms Mach number 
of the flow ({\em Padoan et al.}, 1997; {\em Nordlund and Padoan}, 1999;
{\em Ostriker et al.}, 1999; {\em Li et al.}, 2004). v) If the kinetic 
energy exceeds the magnetic energy, the distribution of the magnetic field 
strength, $B$, is very intermittent and is correlated with the gas density, $n$. 
The scatter plot of $B$ versus $n$ shows a very large dispersion and a well 
defined power law upper envelope ({\em Padoan and Nordlund}, 1999; 
{\em Ostriker et al.}, 2001). If the kinetic energy is 
comparable to the magnetic energy, strong density enhancements are still 
possible in the direction of the magnetic field, but fluctuations of $B$ 
are small and independent of $n$. vi) The flow velocity is correlated to 
the gas density. Because density is increased by shocks, where the velocity 
is dissipated, dense filaments and cores have lower velocity than the low 
density gas ({\em Padoan et al.}, 2001b). vii) The mass distribution of gravitationally 
unstable turbulent density peaks is very close to the stellar IMF and follows
the analytical model of {\em Padoan and Nordlund} (2002) ({\em Li et al.}, 2004; 
{\em Tilley and Pudritz}, 2004).

There is now substantial observational evidence indicating that these main
properties of supersonic MHD turbulence are indeed found in molecular clouds.
The comparison of numerical simulations of turbulence
with observational data was pioneered by {\em Falgarone et al.} (1994) and 
continued by many others (e.g., {\em Padoan et al.} 1998, 1999; {\em Padoan et al.} 
2001a,b; {\em Ostriker et al.}, 2001; {\em Ballesteros-Paredes and Mac Low}, 2002; 
{\em Ossenkopf}, 2002; {\em Padoan et al.}, 2004a; {\em Gammie et al.}, 2003; 
{\em Klessen et al.}, 2005; {\em Esquivel and Lazarian}, 2005). 

However, with the exception of several works by {\em Padoan et al.} and the work 
of {\em Ossenkopf} (2002), where post-processed three dimensional non-LTE radiative 
transfer calculations were carried out, all other studies are based on a 
simple comparison of density and velocity fields in the simulations with the observed 
quantities. Some recent works addressing the comparison of turbulence simulations
and observational data include studies of velocity scaling, showing that 
molecular cloud turbulence is driven on large scale (e.g., {\em Ossenkopf and Mac Low}, 
2002; {\em Heyer and Brunt}, 2004) and studies of core properties, showing that turbulent 
flows naturally generate dense cores with shapes, internal turbulence, rotation 
velocity and magnetic field strength consistent with the observations 
(e.g., {\em Padoan and Nordlund}, 1999; {\em Gammie et al.}, 2003; 
{\em Tilley and Pudritz}, 2004; {\em Tilley and Pudritz}, in preparation; 
{\em Li et al.}, 2004).

\bigskip
\noindent
\textbf{5.2. Collapse and Fragmentation of Molecular Cloud Cores into Low-Mass Stars}
\bigskip

Over the past several years two dominant models 
of how stars acquire their final mass have emerged:
Direct Gravitational Collapse and Competitive Accretion. In both theories, 
a star initially forms when gravitational bound gas collapses. In the 
gravitational collapse scenario, after a protostar has consumed or 
expelled all the gas in its initial core, it may continue accreting 
from the parent clump, but it will not significantly alter its mass 
({\em McKee and Tan}, 2003; {\em Padoan et al.}, 2005; {\em Krumholz et al.}, 2005b). 
Competitive accretion, in contrast, requires that the amount accreted after 
consuming the initial core be substantially larger than the protostellar mass. 

{\em Krumholz et al.} (2005a) define 
$f_m  \equiv  \dot{m}_{*}t_{dyn}/{m}_{*}$   
as the fractional change in mass that a protostar of mass $m_{*}$ 
undergoes each dynamical time $t_{dyn}$ of its parent clump, starting 
after the initial core has been consumed by the accreting protostar.
Gravitational collapse theory suggests that $f_m \ll 1$, while competitive 
accretion {\em requires} $f_m \gg 1$.  In recent work examining the 
plausibility of competitive accretion, {\em Krumholz et al.} (2005a) considered 
two possible scenarios. In the first scenario, the gas the protostar is accreting 
is not accumulated into bound structures on scales smaller than the entire 
clump. For unbound gas, self-gravity may be neglected and the entire problem 
can be treated as Bondi-Hoyle accretion in a turbulent medium of 
non-self-gravitating gas onto a point particle. In a companion paper
({\em Krumholz et al.}, 2006) they develop the theory for Bondi-Hoyle
accretion in a turbulent medium. Using this theory they 
derive the accretion rate for such a turbulent medium and they confirm 
their theory with detailed, high resolution, converged AMR simulations.  
By using this accretion rate and the definition of the virial parameter, 
$\alpha\mathrm{_{vir}} \equiv M\mathrm{_{vir}}/M$, and the dynamical time,
$t\mathrm{_{dyn}} \equiv R/\sigma$, where $\sigma$ is the velocity dispersion 
in the gas, they show that the accretion of unbound gas gives 
$f\mathrm{_{m-BH}} = (14.4,3.08\frac{L}{R}) \phi\mathrm{_{BH}} 
\alpha\mathrm{_{vir}} \! ^{-2} (\frac{m_{*}}{M})$ for a (spherical, 
filamentary) star-forming region, where $\phi\mathrm{_{BH}}$ represents the
effects of turbulence ({\em Krumholz et al.}, 2005a). 

From this it is clear that competitive accretion is most effective in 
low mass clumps with virial parameters $\alpha\mathrm{_{vir}} \ll 1$. 
They then examined the observed properties of a large range of star forming 
regions spanning both low mass and high mass stars and computed the properties 
for each region yielding $\alpha\mathrm{_{vir}}$, $\phi\mathrm{_{BH}}$ and 
$f\mathrm{_{m-BH}}$. In virtually every region examined, the virial 
parameter $\alpha\mathrm{_{vir}} \sim 1$ and $f\mathrm{_{m-BH}} \ll1$. 
Thus {\em none} of the star-forming regions are consistent with 
competitive accretion, but all are consistent with direct gravitational collapse.
{\em Edgar and Clarke}, (2004) examined Bondi-Hoyle accretion onto stars
including radiation pressure effects and found that radiation pressure
halts further accretion around stars more massive than $\sim 10 M\odot$.  
It is important to point out that while this result may be true for
accretion of unbound gas onto a point particle, it is not true
for global collapse and accretion from a bound core  
as shown in more realistic full 3D radiation hydrodynamics simulations by
{\em Krumholz et al.} (2005c), who form massive stars with $M \sim 40 M\odot$.
If {\em Edgar and Clarke's} results do hold for Bondi-Hoyle accretion (but not
accretion from a core), then the only way for massive stars to grow in competitive 
accretion is by direct collisions. This requires densities of $10^8$ $\mathrm {pc^{-3}}$,
$\sim$ 3 orders of magnitude larger than any observed in the galactic plane. 
Furthermore, no competitive accretion model to date has included 
the effects of radiation pressure; a glaring omission if the model is attempting to 
explain high mass stars. It follows that competitive accretion is 
not a viable mechanism for producing the stellar IMF. 

In a second possible competitive accretion scenario, {\em Krumholz et al.} 
(2005a, supplemental section) examined another way that a star could increase its mass
by capturing and accreting other gravitationally bound cores. Their
theory results in the calculation of $f\mathrm{_{m-cap}}$, the fractional change 
in mass that a protostar undergoes by capturing bound cores. As found with 
$f\mathrm{_{m-BH}}$, all the values are estimated to be three more orders of 
magnitude below unity and again, competitive accretion is found not to work.  
  
If competitive accretion is clearly not supported by observations 
in any known star forming region, why do the simulations ({\em Bonnell et al.},1998, 2001; 
{\em Bate et al.}, 2005) almost invariably find competitive accretion 
to work? Is there a fundamental flaw in the methodology used in 
competitive accretion scenarios (SPH) or is the problem one of physics 
and initial conditions? As {\em Krumholz et al.} (2005a) point out, 
all competitive accretion have virial parameters $\alpha\mathrm{_{vir}} \ll 1$.  
Some of the simulations start with $\alpha\mathrm{_{vir}} \approx 0.01$ 
as a typical choice ({Bonnell et al.}, 2001a,b;
{\em Klessen and Burkert}, 2000, 2001). For 
other simulations the virial parameter is initially of order unity but 
decreases to $\ll1$ in a crossing time as turbulence decays 
({\em Bonnell et al.}, 2004; {\em Bate et al.}, 2002a,b, 2003). 
It is also noteworthy that many of the 
simulations begin with clumps of mass considerably smaller ($M \leq 100 M_\odot$)
than that typically one found in star forming regions $\sim 5000M\odot$ 
({\em Plume et al.}, 1997).  {\em Krumholz et al.} 
(2005a) show that for competitive accretion to work, 
$\alpha\mathrm{_{vir}} \! ^2 M < 10 M_\odot$, but for typical star
forming regions 
$\alpha\mathrm{_{vir}} \approx 1$ and M $\approx 10^2$--$10^4 M_{\odot}$
and the inequality is almost never satisfied. 

One reason why simulations evolve to $\alpha\mathrm{_{vir}} \ll 1$ is that they 
omit feedback from star formation. Observations by {\em Quillen et al.} 
(2005) show that outflows inject enough energy on the scale of a clump
to sustain turbulence 
thereby keeping the virial parameter from declining to values much less 
than unity. Another possible reason is that the simulations consider isolated 
clumps with too little material. Real clumps are embedded in larger molecular 
clouds where larger scale turbulent motions can cascade down to the 
clump scale preventing the rapid decay of the turbulence.

\bigskip
\noindent
\textbf{5.3. 3D Collapse with Radiation}
\bigskip

Radiation transfer is an important element in both low mass and high mass
star formation. {\em Boss et al.} (2000) carried out the first 3D simulations 
including radiation transfer to study the effect of radiation on the formation
of stars in collapsing molecular cloud cores. Starting from cores with Gaussian
initial density profiles, this work compared collapse calculations based on the
isothermal and barotropic approximations with the more realistic case of detailed 
radiative transfer in the Eddington approximation. The use of the isothermal equation of
state resulted in a collapse leading to a thin isothermal filament ({\em Truelove et al.}, 
1997). In the more realistic case with nonisothermal heating using radiative
transfer in the Eddington approximation, they showed that thermal retardation
of the collapse caused the formation of a binary protostar system at the same
maximum density where the isothermal collapse yielded a thin filament. Eventually, 
the binary clumps evolved into a central protostar surrounded by spiral arms containing 
two more clumps. The corresponding collapse using the barotropic approximation 
allowed a transition from an isothermal optically thin to an optically thick flow.
It resulted in a transient binary merging into a central object surrounded by 
spiral arms with no evidence of further fragmentation.  
The barotropic result differs significantly from the Eddington result at the same 
maximum density indicating the importance of detailed radiative transfer effects.  

\begin{figure}[h]
\centering
\includegraphics[width=8cm]{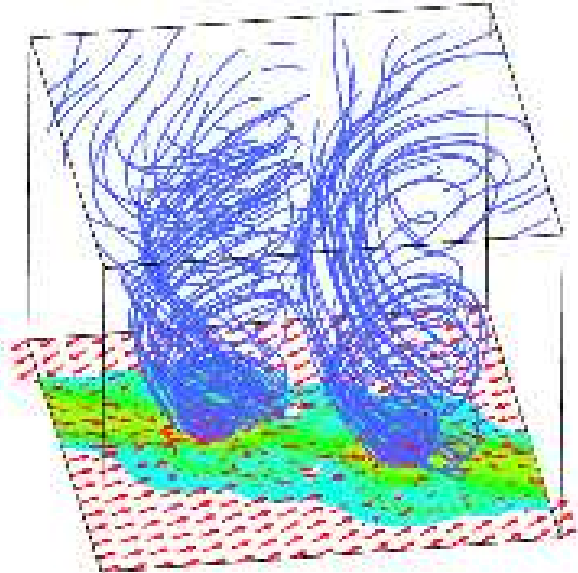}
\caption{Binary outflows. Magnetic field lines, velocity field vectors
and density distribution on the $z=0$ plane. Taken from model DL of 
{\em Machida et al} (2005b). 
\label{Fig:Machida2005b}}
\end{figure}

{\em Boss et al.} (2000) examined the differences in the use of a barotropic 
stiffened EOS approximation and radiative transfer. In the former, the thermal 
properties of the gas are specified solely as a function of density. This
implies that a single global value of a critical density represents the entire 
calculation and its value depends weakly on the assumed geometry of the cloud 
({\em Inutsuka and Miyama}, 1997). In 3-D the effective value of this critical 
density depends on the local geometry surrounding a fluid element. In addition, 
whereas the specific entropy of a gas parcel generally depends on the thermal history
of the parcel, the specific entropy of the gas using the barotropic approximation 
depends solely on the density. Thus the derived pressures used in the momentum 
equations will differ between a calculation using a stiffened EOS approximation 
and a fully consistent calculation using radiative transfer. As a result, the 
temperature is determined not simply by adiabatic compression, but by compressional 
heating in a 3D volume with highly variable optical depth. Thus the
dependence of the temperature on the density cannot be represented with a simple 
barotropic approximation with any great accuracy. This causes concern for the validity 
of current  simulations of multiple star formation and cluster formation, since 
essentially all use either the isothermal or the barotropic approximation. Recent 
work by {\em Whitehouse and Bate} (2005) examined core collapse with radiation
and the adequacy of the barotropic approximation as well.

\bigskip
\noindent
\textbf{5.4. The Debate Over Disk Fragmentation}
\bigskip

As pointed out in section 2, most simulations to date, performed with
SPH, find circumstellar disk fragmentation and rapid ejection of BDs in
most cases ({\em Bate et al.}, 2003;{\em Delgato-Donate et al.},
2004; {\em Goodwin et al.}, 2004a,b) even possibly to an excess of BD 
and low-mass companions ({\em Goodwin et al.}, 2004b). Furthermore, 
most simulations are terminated at an arbitrary time, when much of the gas is still
present; hence the simulations may provide only a lower bound to the 
number of companions that would be produced. As discussed in section 2, 
these simulations are increasingly contradicted by recent observations.  
{\em Goodwin and Kroupa} (2005) have recently pointed out that
observations suggest that individual cores produce at most 2-3 protostars.
A possible reason for this problem is the absence of a magnetic field 
and inaccurate thermodynamics (i.e. barotropic equation rather than
radiative transfer) in the SPH simulations. 

However, there is an alternative explanation: 
The SPH simulations are likely not converged. Indeed, recent high
resolution AMR simulations of the collapse and fragmentation of
turbulent cores ({\em Klein et al.}, 2004a,b) show that a magnetic field
and radiative transfer is not required to explain the observational results of 
{\em Goodwin and Kroupa} (2005). {\em Klein et al.}, (2004a,b) find that 
over a broad range of turbulent Mach numbers ($M \sim 1-3$) and 
rotational energies ($\beta \sim 10^{-4}-10^{-1}$) low order single 
or binary stars are formed through fragmentation of the
core, not the ensuing disk. Is it then a matter of faith in one numerical
method or the other? SPH versus AMR? Not really, it is rather a matter
of testing the convergence of the numerical simulations, irrespective of
the method. 

{\em Fisher et al.} (2006, in preparation) 
have performed high resolution, full convergence studies using the 
SPH code Gadget with the same model and initial conditions as 
{\em Goodwin et al.} (2004a), except for the initial turbulent seed.  
They confirmed the results of {\em Goodwin et al.}, (2004a) at the same low resolution 
(only one smoothing kernel per minimum Jeans mass). But they have also 
found {\em no convergence} of either the multiple number of companion 
protostars produced or the time for multiple fragmentation to occur in 
the disk, with up to 32 times the resolution of {\em Goodwin et al.}, 
(2004a). This suggests that the current SPH simulations showing high 
order multiple disk fragmentation may be grossly under resolved. If so,
the disagreement with the observations may not be due only to the absence
of the magnetic field and radiation, but also to unconverged numerics.
Recent grid-based simulations attempting to study fragmentation in isolated disks
(cf. {\em Duriesen} this book) indicate that for 2D axisymmetric disk studies 
very high resolution, 256 cells in the radial direction alone
({\em Ostriker} private communication), is required to demonstrate that disks do
not suffer from numerical instability. This level of resolution is far
beyond any SPH simulation of cores or clusters to date showing disk fragmentation,
and would be difficult to achieve also with AMR simulations 
starting from globally collapsing cores.

\begin{figure}[h]
\centering
\includegraphics[width=8cm]{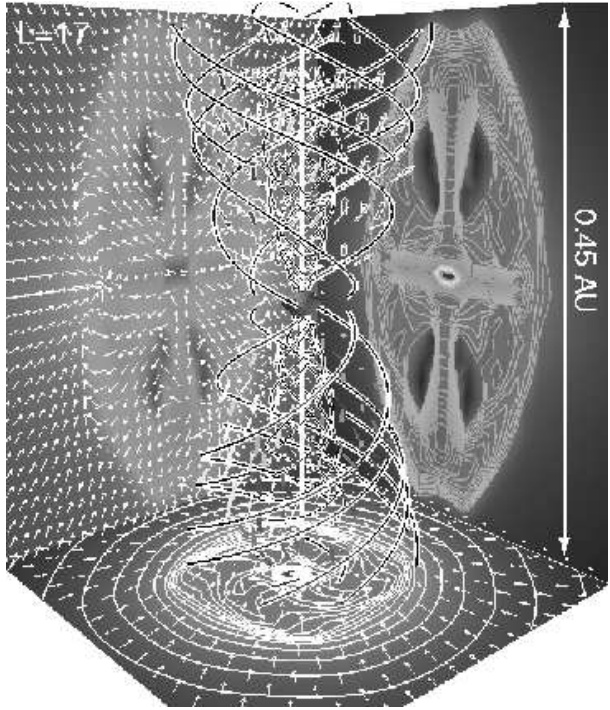}
\caption{Bird's-eye view of the magnetic field lines in the fast 
         jet emanating from the central region around a protostar. 
         Only the 17th grid in the nested-grid resistive MHD 
         calculations by {\it Machida et al.} (2006, in preparation) 
         is shown. The density contours and velocity vectors are projected 
         on the walls on both sides. 
         \label{fig:2ndCollapse}
        }
\end{figure}

At the time of this conference, the debate between SPH and AMR with 
respect to the issue of disk fragmentation continues, but it is our 
strong opinion that all simulations using SPH or AMR
must demonstrate adequate convergence to be credible. This should apply
also to physical systems that may display chaotic behavior, such as the
gas dynamics in a molecular cloud core. A high sensitivity to initial conditions 
may result in statistical distributions of the measured quantities
(e.g., number of collapsing objects and their formation time) around some 
mean value. In this case, an expensive approach to test numerical convergence
would consist of running a large number of experiments at each resolution
and test for the convergence of the statistical distributions of the 
quantities of interest. A less expensive approach is to measure quantities 
with a weak sensitivity to initial conditions because they already represent 
the average of some statistical distribution, or just because they are even
more sensitive to the numerical resolution than to the initial conditions. For example, 
{\em Fisher et al.} (2006) find that as they increase the resolution 
of their SPH simulations, the time of disk fragmentation (when the first object is formed) 
increases monotonically with resolution, a sign of lack of convergence in the SPH
simulations, rather than a signature of chaotic behavior.

\bigskip
\noindent
\textbf{5.5. Star formation in a cloud with Magnetic Field and Rotation}
\bigskip

\noindent
{\em 5.5.1. Fragmentation of the First Core.}
A non-rotating cloud core without magnetic field contracts in a 
self-similar manner to form a first core that is composed of molecular 
hydrogen ({\it Larson}, 1969; {\it Masunaga et al.}, 1998). 
Recently, nested grid MHD simulations ({\em Machida et al.}, 2004, 2005a) 
have revealed that i) a rotating magnetized core evolves maintaining a 
ratio of angular speed to magnetic field strength at the center
$\Omega_c/B_c\simeq {\rm const}$; and ii) $\Omega_c$ and $B_c$ are well 
correlated at the first core phase and satisfy the ``magnetic flux -- spin relation''  
as $\Omega_c^2/0.2^2 4\pi G \rho_c + B_c^2 / 0.36^2 8\pi c_s^2 \rho_c \simeq 1$,
using a central density $\rho_c$ and isothermal sound speed $c_s$.
This is regarded as an appearance of self-similarity in magnetized 
rotating cores.

The fragmentation of the first core is regarded as one of the mechanisms 
to explain close binary systems ({\em Bonnell and Bate}, 1994b; {\em Bate}, 1998).
The magnetic field affects the rotational motion (magnetic braking) and thus 
the fragmentation. Whether the magnetic field stabilizes the first core 
against the fragmentation or not ({\em Machida et al.}, 2005b; {\em Ziegler}, 2005) 
is attracting attention in relation to the binary formation.
As well as non-magnetic cores (see $\S$3.a.5), a magnetized first core can 
fragment if it is rotating sufficiently fast, 
$\Omega_c \simeq 0.2 (4\pi G \rho_c)^{1/2}$ ({\em Machida et al.}, 2005b) and has
only a weak magnetic field, $B_c \lesssim 0.3 (8\pi c_s^2 \rho_c)^{1/2}$.
Simulations show that increasing the magnetic field strength, fragmentation is 
stabilized by the suppression of rotational motion by magnetic braking
(see also {\em Hosking and Whitworth}, 2004). This is not found by {\em Boss} (2002),
but his model equation is not fully consistent with MHD and does not account for
magnetic braking. In order to achieve enough rotation to cause fragmentation,
the initial $\Omega$-to-$B$ ratio must satisfy the condition 
$(\Omega/B)_{init} > 0.39 G^{1/2}c_s
\sim 1.7 \times 10^{-7}(c_s/0.19\mathrm{km\,s^{-1}})^{-1}\mu \mathrm{G}^{-1}$
({\em Machida et al.}, 2005b).

When $\vec{B}$ and angular momentum $\vec{J}$ are not parallel to each other,
the magnetic braking works more efficiently for the component of $\vec{J}$
perpendicular to $\vec{B}$. A magnetically dominated cloud core whose 
$\Omega$-to-$B$ ratio is less than the above critical value forms a disk 
perpendicular to the magnetic field ({\em Matsumoto and Tomisaka}, 2004)
and an outflow is ejected in the direction of the local magnetic field, 
even if $\vec{B}$ is not parallel to $\vec{J}$.
The difference between the local field in the vicinity of a protostar and on
the cloud scale is restricted to $\lesssim 30\deg$ for this case.
 
If the first core is fragmented into binary or multiple cores,
each fragment spins and multiple outflows (Fig.\ref{Fig:Machida2005b}) are ejected 
({\em Machida et al.}, 2004, 2005b; {\em Ziegler}, 2005; {\em Banerjee and Pudritz}, 2006),
which explains binary outflows ({\em Liseau et al.}, 2005).

\noindent
{\em 5.5.2. Second Collapse with Magnetic Field.}
Once the dissociation of molecular hydrogen starts at the central 
region of the first core, the second collapse begins. 
Further calculation of the evolution to form the second core
(i.e., a protostar) requires the inclusion of resistivity in the MHD description, 
as the high density reduces the degree of ionization and the conductivity of 
the medium. {\it Machida et al.} (2006, in preparation) have adopted 
parametrized resistivity as a function of density in their nested-grid 
resistive MHD code, and extended the calculations of the collapse. 
During the isothermal phase, the magnetic Reynolds number is a decreasing 
function of density. If the magnetic Reynolds number decreases below unity, 
the magnetic field is effectively decoupled from the collapsing gas.  
However, the temperature of the gas becomes sufficiently high ($\sim 10^3$K) 
that the magnetic field is re-coupled again with the collapsing gas. 
This relatively sudden grab of the field lines tends to make 
a very collimated fast outflow around the second core.  
Fig. \ref{fig:2ndCollapse} shows a snapshot of the propagation 
of a fast jet from the protostar. The density distribution on the cross 
section is also shown on the left and right walls. Note that bow shocks are 
clearly seen in the density plot. The results of this calculation indicate that  
a realistic modeling of the evolution of temperature and resistivity as a function
of density is required for a precise description of the jet.

\bigskip
\textbf{6.SUMMARY AND FUTURE DIRECTIONS}
\bigskip

\noindent
\textbf{6.1. Summary}
\bigskip

Observations of molecular clouds and cores provide a wealth of data
that are important constraints for initial conditions of numerical
simulations. Large scale simulations should be consistent
with the turbulent nature of the ISM and the observed properties of 
prestellar cores should emerge self-consistently from the simulations.
All calculations must adhere to the Jeans condition for grid-based 
schemes or a well established equivalent for particle-based schemes.  
It is important to stress that the Jeans condition is a necessary 
but not sufficient condition to guarantee avoidance of artificial 
fragmentation. Simulations must be tested at more resolved Jeans 
numbers to establish convergence. 

The importance of turbulence in star formation is now well accepted.
A very large spatial resolution is required to simulate the turbulent
fragmentation, barely achieved by the largest grid-based simulations.
Present SPH simulations fall several orders of magnitude below the 
required spatial resolution. It is possible that almost no available
simulations have yet accurately tested the effect of turbulent 
fragmentation.

Magnetic fields play a crucial role in the star formation process.  At
this time there are no 3D, self-gravitational MHD simulations that have
evolved stars from turbulent clouds. Grid based schemes such as AMR have
developed high order accurate PPM and Godunov based MHD that can provide
accurate solutions across several orders of magnitude of collapse. SPH
has developed cruder approaches to MHD that appear to be rather poor even
for very mild shocks, but this will hopefully improve. Godunov approaches
to MHD for SPH may provide increased accuracy. 

Radiation transport has been shown to play a significant role in the 
outcome of fragmentation to binary and multiple systems. It has been 
shown that side stepping the issue of radiation transport with a 
barotropic approximation can lead to incorrect results. Current AMR 
calculations have implemented radiation transfer in the flux-limited 
approximation and have used this for both low mass and high mass star 
formation. Radiation transfer schemes have recently been developed for SPH 
as well.

Current star formation simulations are not yet adequate to accurately span
the many orders of magnitude of density and spatial range necessary to
account for stars from initially turbulent clouds and encompassing all the
relevant physics. In our opinion, AMR approaches with recent developments 
coupling radiation transfer and MHD hold out the best promise for achieving 
that goal. SPH, while making significant strides in the last few years, 
is still faced with difficult challenges of accurate handling of turbulence, 
strong shocks, MHD, and radiation transfer. However, we anticipate further 
progress with SPH.

Current calculations are still not adequate to explain the stellar IMF. 
Of the two dominant theories for the origin of stellar masses described
as direct gravitational collapse and competitive accretion, observations appear
to favor the first. Recent theoretical work by {\em Krumholz et al.}
(2005a) has now demonstrated that seed protostars cannot gain mass
efficiently by competitive accretion processes in observed
star-forming regions that are approximately in virial balance. 
There is no observational evidence for the existence of regions that 
are far from virial balance, as required by competitive accretion models.
This suggests that competitive accretion is not a viable mechanism for
producing the stellar IMF and that current simulations resulting in 
competitive accretion must have initial or boundary conditions 
inconsistent with the observations, or rather neglect some crucial physics.
Theoretical efforts directed toward the picture of direct gravitational collapse 
set up by turbulent fragmentation appear to be more promising.

\bigskip
\noindent
\textbf{6.2. Future Directions for Numerical Simulation}
\bigskip

The lack of demonstrated convergence for most simulations in the field
presents us with uncalibrated results. We strongly emphasize that future
simulations should demonstrate numerical convergence before detailed
comparison with observations can be credible. Otherwise there is no way
to normalize the accuracy of large scale simulations and the results of
such simulations will not advance our understanding of low mass star
formation. Convergence tests should be always carried out irrespective 
of the numerical approach. A better understanding of the
numerical treatment of disk fragmentation must occur to clear up the
current discrepancies between AMR and SPH.

Future simulations of low mass star formation must endeavor to include 
MHD and radiation transfer. With the development of accurate approaches 
to these processes, we can expect to see simulations become more 
relevant to addressing the observations. For simulations to make a real 
connection to the observations, detailed line profiles and continuum
sub-mm and mm maps should be calculated with 3D radiative transfer codes.
Approaches that go beyond the Eddington approximation and flux-limited diffusion 
such as $S_N$ transport and Monte Carlo need further development to work efficiently 
in full 3D simulations. They will become increasingly important in 
treating the flow of radiation in highly inhomogeneous regions. 
As future simulations encompass multi-coupled physics, 
significant progress must be made in algorithms that improve
the parallel scalability. That is necessary in order to simulate 
the full dynamic range of collapse and fragmentation from clouds to stars,
while capturing all the relevant physics.

\bigskip

\centerline\textbf{REFERENCES}
\bigskip
\parskip=0pt
{\small
\baselineskip=11pt

\refs Allen A., Li Z.-Y., and Shu F.~H.\ (2003) {\em Astrophys. J., 599}, 363-379.  
 
\refs Alves J.~F., Lada C.~J., and Lada E.~A.\ (2001) {\em Nature, 409}, 
159-161.  
 
\refs Bacmann A., Andr{\'e} P., Puget J.-L., Abergel A., Bontemps S., 
and Ward-Thompson D.\ (2000) {\em Astron. Astrophys., 361}, 555-580.  
 
\refs Ballesteros-Paredes J., Gazol A., Kim J., Klessen R.~S., Jappsen 
A.-K., and Tejero E.\ (2006) {\em Astrophys. J., 637}, 384-391.  
 
\refs Ballesteros-Paredes J., Klessen R.~S., and V{\'a}zquez-Semadeni E.\ 
(2003) {\em Astrophys. J., 592}, 188-202.  
 
\refs Ballesteros-Paredes J. and Mac Low M.-M.\ (2002) {\em Astrophys. J., 570}, 
734-748.  
 
\refs Balsara D.~S. and Kim J.\ (2004) {\em Astrophys. J., 602}, 1079-1090.  
 
\refs Balsara D.~S. and Spicer D.~S.\ (1999) {\em J. Comp. Phys., 149}, 270-292.  
 
\refs Banerjee R. and Pudritz R.~E.\ (2006) {\em Astrophys. J.}, in press.

\refs Banerjee R., Pudritz R.~E., and Holmes L.\ (2004) {\em Mon. Not. R. Astron. Soc., 355}, 
248-272.  
 
\refs Barrado y Navascu{\'e}s D., Mohanty S., and Jayawardhana R.\ (2004) {\em 
Astrophys. J., 604}, 284-296.  
 
\refs Bastien P., Cha S.-H., and Viau S.\ (2004) {\em Rev. Mex. Astron. Astrofis., 22}, 144-147.  
 
\refs Basu S. and Mouschovias T.~C.\ (1994) {\em Astrophys. J., 432}, 720-741.  
 
\refs Bate M.~R.\ (1998) {\em Astrophys. J., 508}, L95-L98.  
 
\refs Bate M.~R. and Bonnell I.~A.\ (2005) {\em Mon. Not. R. Astron. Soc., 356}, 1201-1221.  
 
\refs Bate M.~R., Bonnell I.~A., and Price N.~M.\ (1995) {\em Mon. Not. R. Astron. Soc., 277}, 
362-376. 

\refs Bate M.~R., Bonnell I.~A., and Bromm V.\ (2002a) {\em Mon. Not. R. Astron. Soc., 332}, 
L65-L68.  
 
\refs Bate M.~R., Bonnell I.~A., and Bromm V.\ (2002b) {\em Mon. Not. R. Astron. Soc., 336}, 
705-713.  
 
\refs Bate M.~R., Bonnell I.~A., and Bromm V.\ (2003) {\em Mon. Not. R. Astron. Soc., 339}, 
577-599.   
 
\refs Beresnyak A., Lazarian A., and Cho J.\ (2005) {\em Astrophys. J., 624}, L93-L96.  
 
\refs Berger M.~J. and Colella P.\ (1989) {\em J. Comp. Phys., 82}, 64-84.  
 
\refs Blandford R.~D. and Payne D.~G.\ (1982) {\em Mon. Not. R. Astron. Soc., 199}, 883-903.  
 
\refs Bodenheimer P., Burkert A., Klein R.~I., and Boss A.~P.\ (2000) In {\em Protostars 
and Planets IV} (V. Mannings et al., eds.), pp. 675-690. Univ. of Arizona, Tucson. 
 
\refs Boldyrev S.\ (2002) {\em Astrophys. J., 569}, 841-845.  
 
\refs Boldyrev S., Nordlund {\AA}., and Padoan P.\ (2002a) {\em Astrophys. J., 573}, 
678-684.  

\refs Boldyrev S., Nordlund {\AA}., and Padoan P.\ (2002b) {\em Phys. Rev. L., 89}, 
031102-031105. 
 
\refs Bonnell I.~A.\ (1994) {\em Mon. Not. R. Astron. Soc., 269}, 837-848.  
 
\refs Bonnell I.~A. and Bate M.~R.\ (1994a) {\em Mon. Not. R. Astron. Soc., 269}, L45-L48.  
 
\refs Bonnell I.~A. and Bate M.~R.\ (1994b) {\em Mon. Not. R. Astron. Soc., 271}, 999-1004.  

\refs Bonnell I.~A., Bate M.~R., and Zinnecker H.\ (1998) {\em Mon. Not. R. Astron. Soc., 298}, 
93-102.
 
\refs Bonnell I.~A., Bate M.~R., Clarke C.~J., and Pringle J.~E.\ (2001a) {\em 
Mon. Not. R. Astron. Soc., 323}, 785-794.  
 
\refs Bonnell I.~A., Clarke C.~J., Bate M.~R., and Pringle J.~E.\ (2001b) {\em 
Mon. Not. R. Astron. Soc., 324}, 573-579.  
 
\refs Bonnell I.~A., Bate M.~R., and Vine S.~G.\ (2003) {\em Mon. Not. R. Astron. Soc., 343}, 
413-418.    
 
\refs Bonnell I.~A., Vine S.~G., and Bate M.~R.\ (2004) {\em Mon. Not. R. Astron. Soc., 349}, 
735-741.  
 
\refs Bonnor W.~B.\ (1956) {\em Mon. Not. R. Astron. Soc., 116}, 351-359. 
 
\refs Bonnor W.~B.\ (1957) {\em Mon. Not. R. Astron. Soc., 117}, 104-117. 
 
\refs Boss A.~P.\ (1981) {\em Astrophys. J., 250}, 636-644.  
 
\refs Boss A.~P.\ (2002) {\em Astrophys. J., 568}, 743-753.  
 
\refs Boss A.~P. and Myhill E.~A.\ (1992) {\em Astrophys. J. Suppl., 83}, 311-327.  

\refs Boss A.~P., Fisher R.~T., Klein R.~I., and McKee C.~F.\ (2000) {\em Astrophys. J., 
528}, 325-335.   
 
\refs Bourke T.~L., Myers P.~C., Robinson G., and Hyland A.~R.\ (2001) {\em 
Astrophys. J., 554}, 916-932.  
 
\refs Brio M. and Wu C.~C.\ (1988) {\em J. Comp. Phys., 75}, 400-422.  
 
\refs Burkert A., Bate M.~R., and Bodenheimer P.\ (1997) {\em Mon. Not. R. Astron. Soc., 289}, 
497-504.  
 
\refs Byleveld S.~E. and Pongracic H.\ (1996) {\em Publ. Astron. Soc. Pac., 13}, 71-74.  
 
\refs Castor J.~I.\ (2004) {\em Radiation Hydrodynamics}. Cambridge Univ. Press.  
 
\refs Cha S.-H. and Whitworth A.~P.\ (2003a) {\em Mon. Not. R. Astron. Soc., 340}, 73-90.  
 
\refs Cha S.-H. and Whitworth A.~P.\ (2003b) {\em Mon. Not. R. Astron. Soc., 340}, 91-104.  
 
\refs Chabrier G.\ (2003) {\em  Publ. Astron. Soc. Pac., 115}, 763-795.  
 
\refs Cho J., Lazarian A., and Vishniac E.~T.\ (2003) {\em Astrophys. J., 595}, 812-823.  
 
\refs Colella P. and Woodward P.~R.\ (1984) {\em J. Comp. Phys., 54}, 174-201.  
 
\refs Crockett R.~K., Colella P., Fisher R.~T., Klein R.~I., and McKee C.~F.\ 
(2005) {\em J. Comp. Phys., 203}, 422-448.  
 
\refs Crutcher R.~M.\ (1999) {\em Astrophys. J., 520}, 706-713.  

\refs Crutcher R.~M., Troland T.~H., Goodman A.~A., Heiles C., Kazes I., 
and Myers P.~C.\ (1993) {\em Astrophys. J., 407}, 175-184.  
 
\refs Crutcher R.~M., Troland T.~H., Lazareff B., Paubert G., and Kaz{\`e}s I.\ 
(1999) {\em Astrophys. J., 514}, L121-L124.  
 
\refs Dai W. and Woodward P.~R.\ (1994) {\em J. Comp. Phys., 115}, 485-514.  
 
\refs Dedner A., Kemm F., Kr{\"o}ner D., Munz C.-D., Schnitzer T., 
and Wesenberg M.\ (2002) {\em J. Comp. Phys., 175}, 645-673.  
 
\refs Delgado-Donate E.~J., Clarke C.~J., and Bate M.~R.\ (2004) {\em Mon. Not. R. Astron. Soc., 
347}, 759-770.  
 
\refs Dimotakis P.~E.\ (2000) {\em J. Fluid. Mech., 409}, 69-98.  
 
\refs Dobler W., Haugen N.~E., Yousef T.~A., and Brandenburg A.\ (2003) {\em Phys. Rev. E, 68}, 026304-026311 
 
\refs Dubrulle B.\ (1994) {\em Phys. Rev. L., 73}, 959-962.  
 
\refs Duch{\^e}ne G., Bouvier J., and Simon T.\ (1999) {\em Astron. Astrophys., 343}, 
831-840.  

\refs Duch{\^e}ne G., Bouvier J., Bontemps S., Andr{\'e} P., and Motte F.\ 
(2004) {\em Astron. Astrophys., 427}, 651-665.  
 
\refs Dykema P.~G., Klein R.~I., and Castor J.~I.\ (1996) {\em Astrophys. J., 457}, 892-921 
 
\refs Ebert R.\ (1957) {\em Zeitschrift Astrophys., 42}, 263-272 
 
\refs Edgar R. and Clarke C.\ (2004) {\em Mon. Not. R. Astron. Soc., 349}, 678-686.  
 
\refs Elmegreen B.~G. and Falgarone E.\ (1996) {\em Astrophys. J., 471}, 816-821 
 
\refs Esquivel A. and Lazarian A.\ (2005) {\em Astrophys. J., 631}, 320-350.  
 
\refs Evans C.~R. and Hawley J.~F.\ (1988) {\em Astrophys. J., 332}, 659-677.  
 
\refs Falgarone E., Puget J.-L., and Perault M.\ (1992) {\em Astron. Astrophys., 257}, 
715-730.  
 
\refs Falgarone E., Lis D.~C., Phillips T.~G., Pouquet A., Porter D.~H., 
and Woodward P.~R.\ (1994) {\em Astrophys. J., 436}, 728-740.  
 
\refs Falkovich G.\ (1994) {\em Phys. Fluids, 6}, 1411-1414.  
 
\refs Fuller G.~A. and Myers P.~C.\ (1992) {\em Astrophys. J., 384}, 523-527.  
 
\refs Galli D., Shu F.~H., Laughlin G., and Lizano S.\ (2001) {\em Astrophys. J., 551}, 
367-386.  
 
\refs Gammie C.~F., Lin Y.-T., Stone J.~M., and Ostriker E.~C.\ (2003) {\em 
Astrophys. J., 592}, 203-216.  
 
\refs Goodman A.~A., Benson P.~J., Fuller G.~A., and Myers P.~C.\ (1993) {\em 
Astrophys. J., 406}, 528-547.  
 
\refs Goodwin S.~P. and Kroupa P.\ (2005) {\em Astron. Astrophys., 439}, 565-569.  
 
\refs Goodwin S.~P., Whitworth A.~P., and Ward-Thompson D.\ (2004a) {\em Astron. Astrophys., 
414}, 633-650.  
 
\refs Goodwin S.~P., Whitworth A.~P., and Ward-Thompson D.\ (2004b) {\em Astron. Astrophys., 
423}, 169-182.  
 
\refs Gudiksen B.~V. and Nordlund {\AA}.\ (2005) {\em Astrophys. J., 618}, 1020-1030.  
 
\refs Haugen N.~E. and Brandenburg A.\ (2004) {\em Phys. Rev. E, 70}, 026405-026411. 
 
\refs Heiles C., Goodman A.~A., McKee C.~F., and Zweibel E.~G.\ (1993) In 
{\it Protostars and Planets III} (E. H. Levy and J. I. Lunine, eds.), pp. 279-326. 
Univ. of Arizona, Tucson.
 
\refs Heyer M.~H. and Brunt C.~M.\ (2004) {\em Astrophys. J., 615}, L45-L48.  
 
\refs Hosking J.~G. and Whitworth A.~P.\ (2004) {\em Mon. Not. R. Astron. Soc., 347}, 1001-1010.  
 
\refs Imamura J.~N., Durisen R.~H., and Pickett B.~K.\ (2000) {\em Astrophys. J., 528}, 
946-964.  
 
\refs Inutsuka S.-I. and Miyama S.~M.\ (1997) {\em Astrophys. J., 480}, 681-693. 
 
\refs Jayawardhana R., Mohanty S., and Basri G.\ (2002) {\em Astrophys. J., 578}, L141-L144.  
 
\refs Jeans J.~H.\ (1902) {\em Phil. Trans. A, 199}, 1-53. 
 
\refs Jijina J., Myers P.~C., and Adams F.~C.\ (1999) {\em Astrophys. J. Suppl., 125}, 161-236.
 
\refs Juvela M., Padoan P., and Nordlund {\AA}.\ (2001) {\em Astrophys. J., 563}, 
853-866.  
 
\refs Kitsionas S. and Whitworth A.~P.\ (2002) {\em Mon. Not. R. Astron. Soc., 330}, 129-136.  

\refs Klein R.~I.\ (1999), {\em Journ. Comp. and Appled Math., 109}, 123-152.

\refs Klein R.~I., Colella P., and McKee C.~F.\ (1990) {\em Publ. Astron. Soc. Pac., 12}, 117-136. 

\refs Klein R.~I., Fisher R.~T., McKee C.~F., and Truelove J.~K.\ (1999) {\em 
numa.conf}, 131-140 

\refs Klein R.~I., Fisher R., and McKee C.~F.\ (2001) In {\em The Formation of Binary Stars} 
(H. Zinnecker and R. D. Mathieu, eds.), pp. 361-370. 
 
\refs Klein R.~I., Fisher R.~T., Krumholz M.~R., and McKee C.~F.\ (2003) {\em 
Rev. Mex. Astron. Astrofis., 15}, 92-96.  

\refs Klein R.~I., Fisher R.~T., McKee C.~F., and Krumholz M.~R.\ (2004a) 
{\em Publ. Astron. Soc. Pac., 323}, 227-234 

\refs Klein R.~I., Fisher R., and McKee C.~F.\ (2004b) {\em Rev. Mex. Astron. Astrofis., 22}, 3-7.  
 
\refs Klein R.~I., Fisher R.~T., McKee C.~F., and Krumholz M.~R.\ (2004c),
In {\em Adaptive Mesh Refinement} (Tomasz Plewa, ed.), pp. 112-118. Chicago University Press
 
\refs Klessen R.~S.\ (2001) {\em Astrophys. J., 556}, 837-846.  
 
\refs Klessen R.~S. and Burkert A.\ (2000) {\em Astrophys. J. Suppl., 128}, 287-319.  
 
\refs Klessen R.~S. and Burkert A.\ (2001) {\em Astrophys. J., 549}, 386-401.  
 
\refs Klessen R.~S., Ballesteros-Paredes J., V{\'a}zquez-Semadeni E., 
and Dur{\'a}n-Rojas C.\ (2005) {\em Astrophys. J., 620}, 786-794.  

\refs Koyama H. and Inutsuka S.-I.\ (2000) {\em Astrophys. J., 532}, 980-993.  
 
\refs Kritsuk A.~G., Norman M.~L., and Padoan P.\ (2006) {\em Astrophys. J., 638}, 
L25-L28.  
 
\refs Krumholz M.~R., McKee C.~F., and Klein R.~I.\ (2004) {\em Astrophys. J., 611}, 
399-412.  

\refs Krumholz M.~R., McKee C.~F., and Klein R.~I.\ (2005a) {\em Nature, 438}, 
332-334.  

\refs Krumholz M.~R., McKee C.~F., and Klein R.~I.\ (2005b) {\em Astrophys. J., 618}, 
757-768.  

\refs Krumholz M.~R., Klein R.~I., and McKee C.~F.\ (2005c) In {\em Massive star birth: 
A crossroads of Astrophysics} (R. Cesaroni et al., eds.), pp. 231-236. Cambridge University Press.  
 
\refs Krumholz M.~R., McKee C.~F., and Klein R.~I.\ (2005d) {\em Astrophys. J., 618}, 
L33-L36.  
 
\refs Krumholz M.~R., McKee C.~F., and Klein R.~I.\ (2006) {\em Astrophys. J., 638}, 
369-381.  
 
\refs Lada C.~J. and Lada E.~A.\ (2003) {\em Ann. Rev. Astron. Astrophys., 41}, 57-115.  
 
\refs Larson R.~B.\ (1969) {\em Mon. Not. R. Astron. Soc., 145}, 271-295. 

\refs Larson R.~B.\ (1981) {\em Mon. Not. R. Astron. Soc., 194}, 809-826.  
 
\refs Li P.~S., Norman M.~L., Mac Low M.-M., and Heitsch F.\ (2004) 
{\em Astrophys. J., 605}, 800-818.  
 
\refs Liseau R., Fridlund C.~V.~M., and Larsson B.\ (2005) {\em Astrophys. J., 619}, 959-967.  
 
\refs Mac Low M.-M. and Klessen R.~S.\ (2004) {\em Rev. Mod. Phys., 76}, 125-194.  
 
\refs Mac Low M.-M., Smith M.~D., Klessen R.~S., and Burkert A.\ (1998) {\em 
Astron. Astrophys. Suppl., 261}, 195-196.  
 
\refs Machida M.~N., Tomisaka K., and Matsumoto T.\ (2004) {\em Mon. Not. R. Astron. Soc., 348}, 
L1-L5.  

\refs Machida M.~N., Matsumoto T., Tomisaka K., and Hanawa T.\ (2005a) {\em 
Mon. Not. R. Astron. Soc., 362}, 369-381.  

\refs Machida M.~N., Matsumoto T., Hanawa T., and Tomisaka K.\ (2005b) {\em 
Mon. Not. R. Astron. Soc., 362}, 382-402.  
  
\refs MacNeice P., Olson K.~M., Mobarry C., de Fainchtein R., and Packer C.\ 
(2000) {\em Comp. Phys. Comm., 126}, 330-354.  
 
\refs Masunaga H. and Inutsuka S.-I.\ (1999) {\em Astrophys. J., 510}, 822-827.  
 
\refs Masunaga H. and Inutsuka S.-i.\ (2000a) {\em Astrophys. J., 531}, 350-365.  
 
\refs Masunaga H. and Inutsuka S.-i.\ (2000b) {\em Astrophys. J., 536}, 406-415.  
 
\refs Masunaga H., Miyama S.~M., and Inutsuka S.-I.\ (1998) {\em Astrophys. J., 495}, 346-369. 
 
\refs Matsumoto T. and Hanawa T.\ (2003) {\em Astrophys. J., 595}, 913-934.  
 
\refs Matsumoto T. and Tomisaka K.\ (2004) {\em Astrophys. J., 616}, 266-282.  
 
\refs McKee C.~F., Zweibel E.~G., Goodman A.~A., and Heiles C.\ (1993) In 
{\it Protostars and Planets III} (E. H. Levy and J. I. Lunine, eds.), pp. 327-342. 
Univ. of Arizona, Tucson.
 
\refs McKee C.~F.\ (1989) {\em Astrophys. J., 345}, 782-801.  
 
\refs McKee C.~F. and Tan J.~C.\ (2003) {\em Astrophys. J., 585}, 850-871.  
 
\refs Mihalas D. and Mihalas B.~W.\ (1984) {\em Foundations of radiation hydrodynamics}. 
Oxford University, Oxford.
 
\refs Mohanty S., Jayawardhana R., and Basri G.\ (2005) {\em Astrophys. J., 626}, 
498-522.  
 
\refs Monaghan J.~J.\ (1989) {\em J. Comp. Phys., 82}, 1-15.  
 
\refs Monaghan J.~J.\ (1994) {\em J. Comp. Phys., 110}, 399-406.  
 
\refs Morris J.~P.\ (1996) {\em Publ. Astron. Soc. Aus., 13}, 97-102.  
 
\refs Motte F., Andre P., and Neri R.\ (1998) {\em Astron. Astrophys., 336}, 150-172.  
 
\refs Motte F., Andr{\'e} P., Ward-Thompson D., and Bontemps S.\ (2001) {\em 
Astron. Astrophys., 372}, L41-L44.  
 
\refs Mouschovias T.~C. and Spitzer L.\ (1976) {\em Astrophys. J., 210}, 326-327. 
 
\refs Myers P.~C., Dame T.~M., Thaddeus P., Cohen R.~S., Silverberg R.~F., 
Dwek E., and Hauser M.~G.\ (1986) {\em Astrophys. J., 301}, 398-422.  
 
\refs Nakajima Y. and Hanawa T.\ (1996) {\em Astrophys. J., 467}, 321-333. 
 
\refs Nakano T., Nishi R., and Umebayashi T.\ (2002) {\em Astrophys. J., 573}, 199-214.  
 
\refs Natta A., Testi L., Muzerolle J., Randich S., Comer{\'o}n F., and Persi 
P.\ (2004) {\em Astron. Astrophys., 424}, 603-612.  
 
\refs Nordlund {\AA}. and Padoan P.\ (1999) In {\em Interstellar Turbulence} 
(J. Franco and A. Carraminana, eds.), pp. 218-231. Cambridge University Press. 

\refs Nordlund {\AA}. and Padoan P.\ (2003) In {\em Turbulence and Magnetic Fields 
in Astrophysics} (E. Falgarone, and T. Passot, eds.), pp. 271-298. Springer.  
 
\refs Norman M.~L. and Bryan G.~L.\ (1999) In {\em Numerical Astrophysics} 
(S. M. Miyama et al., eds.), pp. 19-33. Kluwer Academic. 
 
\refs Nutter D.~J., Ward-Thompson D., Crutcher R.~M., and Kirk J.~M.\ (2004) 
{\em Astron. Astrophys. Suppl., 292}, 179-184.  
 
\refs Ossenkopf V.\ (2002) {\em Astron. Astrophys., 391}, 295-315.  
 
\refs Ossenkopf V. and Mac Low M.-M.\ (2002) {\em Astron. Astrophys., 390}, 307-326.  
 
\refs Ostriker E.~C., Gammie C.~F., and Stone J.~M.\ (1999) {\em Astrophys. J., 513}, 
259-274.  
 
\refs Ostriker E.~C., Stone J.~M., and Gammie C.~F.\ (2001) {\em Astrophys. J., 546}, 
980-1005.  

\refs Padoan P. and Nordlund {\AA}.\ (1997) astro-ph/9706177

\refs Padoan P. and Nordlund {\AA}.\ (1999) {\em Astrophys. J., 526}, 279-294.  
 
\refs Padoan P. and Nordlund {\AA}.\ (2002) {\em Astrophys. J., 576}, 870-879.  
 
\refs Padoan P. and Nordlund {\AA}.\ (2004) {\em Astrophys. J., 617}, 559-564.

\refs Padoan P., Jones B.~J.~T., and Nordlund A.~P.\ (1997) {\em Astrophys. J., 474}, 
730-734. 
 
\refs Padoan P., Juvela M., Bally J., and Nordlund A.\ (1998) {\em Astrophys. J., 504}, 
300-313. 
 
\refs Padoan P., Bally J., Billawala Y., Juvela M., and Nordlund {\AA}.\ (1999) 
{\em Astrophys. J., 525}, 318-329.  
 
\refs Padoan P., Goodman A., Draine B.~T., Juvela M., and Nordlund {\AA}., 
R{\"o}gnvaldsson {\"O}.~E.\ (2001a) {\em Astrophys. J., 559}, 1005-1018.  

\refs Padoan P., Juvela M., Goodman A.~A., and Nordlund {\AA}.\ (2001b) {\em 
Astrophys. J., 553}, 227-234.  
 
\refs Padoan P., Jimenez R., Juvela M., and Nordlund {\AA}.\ (2004a) {\em Astrophys. J., 
604}, L49-L52.  
 
\refs Padoan P., Jimenez R., Nordlund {\AA}., and Boldyrev S.\ (2004b) {\em 
Phys. Rev. L., 92}, 191102-191105. 
 
\refs Padoan P., Kritsuk A., Norman M.~L., and Nordlund {\AA}.\ (2005) {\em 
Astrophys. J., 622}, L61-L64.    
 
\refs Passot T. and V{\'a}zquez-Semadeni E.\ (1998) {\em Phys. Rev. E, 58}, 
4501-4510.  
 
\refs Phillips G.~J. and Monaghan J.~J.\ (1985) {\em Mon. Not. R. Astron. Soc., 216}, 883-895.  
 
\refs Plume R., Jaffe D.~T., Evans N.~J., Martin-Pintado J., and Gomez-Gonzalez 
J.\ (1997) {\em Astrophys. J., 476}, 730-749. 
 
\refs Porter D.~H. and Woodward P.~R.\ (1994) {\em Astrophys. J. Suppl., 93}, 309-349.  
 
\refs Powell K.~G., Roe P.~L., Linde T.~J., Gombosi T.~I., and de Zeeuw D.~L.\ 
(1999) {\em J. Comp. Phys., 154}, 284-309.  
 
\refs Price D.~J. and Monaghan J.~J.\ (2004a) {\em Mon. Not. R. Astron. Soc., 348}, 123-138.  
 
\refs Price D.~J. and Monaghan J.~J.\ (2004b) {\em Mon. Not. R. Astron. Soc., 348}, 139-152.  
 
\refs Price D.~J. and Monaghan J.~J.\ (2004c) {\em Astron. Astrophys. Suppl., 292}, 279-283.  
 
\refs Price D.~J. and Monaghan J.~J.\ (2005) {\em Mon. Not. R. Astron. Soc., 364}, 384-406.  
 
\refs Quillen A.~C., Thorndike S.~L., Cunningham A., Frank A., Gutermuth 
R.~A., Blackman E.~G., Pipher J.~L., and Ridge N.\ (2005) {\em Astrophys. J., 632}, 
941-955.  
 
\refs Reipurth B. and Clarke C.\ (2001) {\em Astron. J., 122}, 432-439.  
 
\refs Ryu D. and Jones T.~W.\ (1995) {\em Astrophys. J., 442}, 228-258.  
 
\refs Ryu D. and Jones T.~W., Frank A.\ (1995) {\em Astrophys. J., 452}, 785-796. 
 
\refs Ryu D., Miniati F., and Jones T.~W., Frank A.\ (1998) {\em Astrophys. J., 509}, 
244-255.  
 
\refs Scalo J., Vazquez-Semadeni E., Chappell D., and Passot T.\ (1998) {\em 
Astrophys. J., 504}, 835-853. 
 
\refs She Z.-S. and Leveque E.\ (1994) {\em Phys. Rev. L., 72}, 336-339.  
 
\refs Shu F.~H., Adams F.~C., and Lizano S.\ (1987) {\em Ann. Rev. Astron. Astrophys., 25}, 23-81.  
 
\refs Shu F.~H., Laughlin G., Lizano S., and Galli D.\ (2000) {\em Astrophys. J., 535}, 
190-210.  
 
\refs Snell R.~L. et al.\ (2000) {\em Astrophys. J., 539}, L101-L105.  
 
\refs Stamatellos D. and Whitworth A.~P.\ (2003) {\em Astron. Astrophys., 407}, 941-955.  
 
\refs Stamatellos D. and Whitworth A.~P.\ (2005) {\em Astron. Astrophys., 439}, 153-158.  
 
\refs Stellingwerf R.~F. and Peterkin R.~E.\ (1994) {\em Mem. Soc. Astron. It., 65}, 991-1011. 
 
\refs Stone J.~M. and Norman M.~L.\ (1992a) {\em Astrophys. J. Suppl., 80}, 753-790.  
 
\refs Stone J.~M. and Norman M.~L.\ (1992b) {\em Astrophys. J. Suppl., 80}, 791-818. 
 
\refs Stone J.~M., Ostriker E.~C., and Gammie C.~F.\ (1998) {\em Astrophys. J., 508}, 
L99-L102.  
 
\refs Sytine I.~V., Porter D.~H., Woodward P.~R., Hodson S.~W., and Winkler 
K.-H.\ (2000) {\em J. Comp. Phys., 158}, 225-238.  
 
\refs T{\'o}th G.\ (2000) {\em J. Comp. Phys., 161}, 605-652.  
 
\refs Tilley D.~A. and Pudritz R.~E.\ (2004) {\em Mon. Not. R. Astron. Soc., 353}, 769-788.  
  
\refs Tomisaka K.\ (1998) {\em Astrophys. J., 502}, L163-167 
 
\refs Tomisaka K.\ (2000) {\em Astrophys. J., 528}, L41-L44.  
 
\refs Tomisaka K.\ (2002) {\em Astrophys. J., 575}, 306-326.  
 
\refs Tomisaka K., Ikeuchi S., and Nakamura T.\ (1988) {\em Astrophys. J., 335}, 239-262.
 
\refs Tomisaka K., Ikeuchi S., and Nakamura T.\ (1989) {\em Astrophys. J., 341}, 220-237. 

\refs Troland T.~H. and Heiles C.\ (1986) {\em Astrophys. J., 301}, 339-345.  
 
\refs Truelove J.~K., Klein R.~I., McKee C.~F., Holliman J.~H., Howell 
L.~H., Greenough J.~A., and Woods D.~T.\ (1998) {\em Astrophys. J., 495}, 821-852. 
 
\refs Tsuribe T. and Inutsuka S.-I.\ (1999a) {\em Astrophys. J., 523}, L155-L158.  
 
\refs Tsuribe T. and Inutsuka S.-I.\ (1999b) {\em Astrophys. J., 526}, 307-313.  
 
\refs van Leer B.\ (1979) {\em J. Comp. Phys., 32}, 101-136.  
 
\refs Vazquez-Semadeni E.\ (1994) {\em Astrophys. J., 423}, 681-692. 
 
\refs Wada K. and Norman C.~A.\ (2001) {\em Astrophys. J., 547}, 172-186.  
 
\refs Whitehouse S.~C. and Bate M.~R.\ (2004) {\em Mon. Not. R. Astron. Soc., 353}, 1078-1094.  
 
\refs Whitehouse S.~C., Bate M.~R., and Monaghan J.~J.\ (2005) {\em Mon. Not. R. Astron. Soc., 
364}, 1367-1377.  
 
\refs Wolfire M.~G., Hollenbach D., McKee C.~F., Tielens A.~G.~G.~M., and Bakes 
E.~L.~O.\ (1995) {\em Astrophys. J., 443}, 152-168.  
 
\refs Yorke H.~W., Bodenheimer P., and Laughlin G.\ (1993) {\em Astrophys. J., 411}, 
274-284.  
 
\refs Yorke H.~W. and Sonnhalter C.\ (2002) {\em Astrophys. J., 569}, 846-862.  
 
\refs Ziegler U.\ (2005) {\em Astron. Astrophys., 435}, 385-395.

\end{document}